\documentclass{amsart}
\usepackage{amssymb}
\usepackage{eucal}
\newtheorem{theorem}{Theorem}[section]
\newtheorem{lemma}[theorem]{Lemma}
\newtheorem{cor}{Corollary}[theorem]
\newtheorem{prop}[theorem]{Proposition}
\theoremstyle{definition}
\newtheorem{definition}{Definition}

\numberwithin{equation}{section}
\begin{document}
\title[Hyperelliptic Kleinian functions]
{Hyperelliptic Kleinian functions and applications}
\author{V M Buchstaber}
\address{National Scientific and
Research Institute of Physico-Technical and Radio-Tech\-ni\-cal
Measurements,  VNIIFTRI\\
Mendeleevo,  Moscow Region,  141570,  Russia}
\author{V Z Enolskii}
\author{D V Leykin}
\address{Theoretical Physics Division\\
NASU Institute of Magnetism\\
36--b Vernadsky str.,   Kiev-680,  252142,  Ukraine}
\date{\today}
\begin{abstract}
We develop the theory of
hyperelliptic Kleinian  functions. As applications we consider
construction of the explicit matrix realization of the hyperelliptic
Kummer varieties,  differential operators to have the hyperelliptic
curve as spectral variety,  solution of the KdV equations by
Kleinian functions.
\end{abstract}
\maketitle
\tableofcontents
\section*{Introduction}

In  this paper we develop  the Kleinian construction of
hyperelliptic Abelian functions,  which is a natural
generalization of the Weierstrass approach in the elliptic
functions theory to the case of a hyperelliptic curve of genus
$g>1$.  Kleinian $ \zeta$ and $ \wp$--functions are defined as
 \[
  \zeta_{i} ( \boldsymbol{ u}) = \frac{ \partial}{ \partial
u_i} \mathrm{ln} \; \sigma ( \boldsymbol{ u}) ,   \quad
  \wp_{ij} ( \boldsymbol{ u}) =- \frac{ \partial^2}{ \partial
u_i \partial u_j}  \mathrm{ln} \; \sigma ( \boldsymbol{ u}) ,   \qquad
i, j=1,  \ldots, g,   \]
where the vector $ \boldsymbol{ u}$ belongs to Jacobian
$ \mathrm{Jac} (V) $ of the
hyperelliptic curve
$V= \{ (y, x)  \in  \mathbb{ C}^2:y^2- \sum_{i=0}^{2g+2}
\lambda_{i}x^i=0 \}$
and the $ \sigma ( \boldsymbol{ u}) $ is the Kleinian $ \sigma$--function.

The systematical study of the $ \sigma$--functions,
which may be related
to the paper of Klein  \cite{kl88},  was an alternative to
the developments of
Weierstrass  \cite{w54a, w54}  (the hyperelliptic generalization of
the Jacobi elliptic functions
$ \mathrm{sn},  \mathrm{cn},  \mathrm{dn}$)  and the purely
$ \theta$--functional theory G{\"o}ppel \cite{go47} and Rosenhain
 \cite{ro51} for genus $2$,  generalized further by Riemann.  The
$ \sigma$ approach was contributed by Burkhardt  \cite{bur88},
Wiltheiss  \cite{wi88},  Bolza  \cite{bo95},  Baker  \cite{ba98} and
others; the detailed bibliography may be found in  \cite{kw15}.  We
would like to cite separately H.F. Baker's monographs
 \cite{ba97, ba07},  worth special attention.

The paper is organized as follows. We recall the basic
facts about hyperelliptic curves in the
Section~ \ref{prelims}. In the Section~ \ref{sigma-defs} we
construct  the explicit expression for the fundamental
$2$--differential of the second kind and derive the solution of
the Jacobi inversion problem in terms of the hyperelliptic
$ \wp$--functions.  We
give in the Section~3 the proof and the analysis of basic
relations for $ \wp$--functions.  It is given  an explicit
description of the $ \mathrm{Jac} (V) $ in $ \mathbb{
C}^{g+ \frac{g (g+1) }{2}}$ as the intersection of cubics. We
introduce coordinates $h_{ij}$  (see below  \eqref{variables}) ,  in terms
of which these cubics are the determinants of
$3 \times3$--matrices,  inheriting in such a way the structure of
Weierstrass  elliptic  cubic. The Kummer variety
$ \mathrm{Kum} (V) = \mathrm{Jac} (V) / \pm$ appears to be the
intersection of  quartics in
$ \mathbb{ C}^{ \frac{g (g+1) }{2}}$ and is described in a whole by
the condition  $  \mathrm{rank} \left ( \{h_{ij}
 \}_{i, j=1,  \ldots, g+2} \right) <4$.
The Section~4 describes some natural applications of the Kleinian
functions theory.

The paper is based on the recent results  partially announced
in  \cite{le95, bel96, el95, be96b}.  The given
results are already used to describe a $2$--dimensional
Schr{\"o}dinger equation  \cite{be95}.

\section{Preliminaries} \label{prelims}

We recall some basic definition from the theory of the
hyperelliptic curves and $ \theta$--functions; see e.g.
 \cite{ba97, ba07, fa73, mu75, gh78, fk80} for the detailed exposition.
 \subsection{Hyperelliptic curves}
The set of points $V (y, x) $ satisfying the
 \begin{eqnarray}
y^2= \sum_{i=0}^{2g+2} \lambda_{i}x^{i}
= \lambda_{2g+2} \prod_{k=1}^{2g+2} (x-e_{k}) =f (x)
 \label{curve}
 \end{eqnarray}
is a model of a plane { \em hyperelliptic curve} of genus $g$,
realized as a $2$--sheeted
covering over Riemann sphere with the { \em branching points}
$e_1,  \ldots, e_{2g+2}$. Any pair  $ (y, x) $ in $V (y, x) $
is called an { \em analytic point}; an analytic point,  which is not
a  branching point is called a { \em regular point}.
 The { \em hyperelliptic involution} $ \phi ( \;) $  (the swap of the
sheets of covering)  acts as $ (y, x)   \mapsto  (-y, x) $,  leaving the
branching points fixed.

To make  $y$ the singlevalued function of $x$ it suffices to
draw $g+1$ cuts,  connecting pairs of branching points
$e_i$---$e_{i'}$ for some partition of $ \{1,  \ldots,  2g+2 \}$ into
the set of $g+1$ disjoint pairs ${i,  i'}$.  Those of $e_j$,  at
which the cuts start we will denote ${a_i}$,  ending points of the
cuts we will denote $b_i$,  respectively; except for one of the
cuts which is denoted by starting point $a$ and ending point $b$.
In the case $ \lambda_{2g+2}  \mapsto 0$ this point $a  \mapsto  \infty$. The
equation of the curve,  in case $ \lambda_{2g+2}=0$ and
$ \lambda_{2g+1}=4$  can be rewritten as  \begin{eqnarray} &y^2=4
P (x) Q (x) ,  \label{alt-curve}  \\ &P (x) = \prod_{i=1}^{g} (x-a_i) ,
 \quad Q (x) = (x-b)  \prod_{i=1}^{g} (x-b_i) . \nonumber
 \end{eqnarray}

The local parametrisation of the point $ (y, x) $ in the  vicinity of
a point $ (w, z) $:   \[ x=z+ \left \{ \begin{array}{ll}  \xi,  \quad
& \text{near regular point} \,  ( \pm w, z) ;  \\  \xi^2, & \text{near
branching point} \,  (0, e_i) ;  \\  \frac{1}{ \xi}, & \text{near regular
point} \,   ( \pm \infty,  \infty) ; \\  \frac{1}{ \xi^2}, & \text{near
branching point} \,   ( \infty,  \infty)    \end{array} \right.   \]
provides the structure of the { \em hyperelliptic Riemann surface}
--- a one-dimensional compact complex manifold. We will employ the
same notation for the plane curve and the Riemann surface ---
$V (y, x) $ or $V$. All curves and Riemann surfaces through the paper
are assumed to be hyperelliptic,  if the converse not stated.

A { \em marking} on $V (y, x) $ is given by the base
point $x_0$ and the canonical basis of cycles
$ (A_1,  \ldots, A_g;B_1,  \ldots, B_g) $ --- the basis in the group of
one-dimensional homologies $H_1 (V (y, x) ,  \mathbb{ Z}) $ on the
surface $V (y, x) $  with the symplectic intersection matrix
$I= \left ( \begin{array}{cc}0&- \mathbf{ 1}_g \\ \mathbf{
1}_g&0 \end{array} \right) $,  where $ \mathbf{ 1}_g$ is the unit
$g \times g$--matrix.

 \subsection{Differentials}
Traditionally three kinds of differential $1$--forms are
distingui \-s \-h \-ed on a Riemann surface.

 \subsubsection{Holomorphic differentials}
or the differentials of the first kind,
are the differential $1$--forms
$ \mathrm{d}u$,  which can be locally given as
$ \mathrm{d}u= ( \sum_{i=0}^ \infty \alpha_i \xi^i) d \xi$ in the vicinity
of any point $ (y, x) $ with some constants $ \alpha_i \in \mathbb{ C}$.
It can be checked directly,  that forms satisfying such a condition
are all of the form $ \sum_{i=0}^{g-1}  \beta_i
x^i \frac{ \mathrm{d}x}{y}$.  Forms $ \{ \mathrm{d}u_i \}_{i=1}^g$,
 \[ \mathrm{d}u_i= \frac{x^{i-1}  \mathrm{d}x}{ y},  \quad
i \in 1,  \ldots, g  \] are the set of { \em canonical
holomorphic differentials} in $H^1 (V,  \mathbb{ C}) $.  The
$g \times g$--matrices of their $A$ and $B$--periods,
 \[2 \omega= \left ( \oint_{A_k} \mathrm{d}u_l \right) ,  \quad
2 \omega'= \left ( \oint_{B_k} \mathrm{d}u_l \right)   \]
are nondegenerate. Under the action of the transformation
$ (2 \omega) ^{-1}$  the vector
$ \mathrm{d} \mathbf{ u}= ( \mathrm{d}u_1,  \ldots,  \mathrm{d}u_g) ^T$ maps
to the vector of normalized holomorphic differentials
$ \mathrm{d} \mathbf{ v}= ( \mathrm{d}v_1,  \ldots,  \mathrm{d}v_g) ^T$
 --- the   vector in
$ H^1 (V,  \mathbb{ C}) $  to satisfy the conditions
$ \oint_{A_k} \mathrm{d}v_k= \delta_{kl}, k, l=1 \ldots, g$. It is known,
that $g \times g$ matrix,
 \[ \tau= \left ( \oint_{B_k} \mathrm{d}v_l \right) = \omega^{-1} \omega' \]
belongs to the { \em upper Siegel halfspace} $ \mathcal{ S}_g$ of degree
$g$,  i.e. it is symmetric  and has a positively defined imaginary
part.

Let us denote by $ \mathrm{Jac} (V) $ the { \em Jacobian} of the curve
$V$,  i.e.  the factor $ \mathbb{ C}^g/ \Gamma$,  where
$ \Gamma=2 \omega \oplus 2 \omega'$ is the lattice generated by the
periods of canonical holomorphic differentials.

{ \em Divisor} $ \mathcal{D}$ is a formal sum of subvarieties of
codimension $1$ with coefficients from $ \mathbb{ Z}$. Divisors on
Riemann surfaces are given by formal sums of analytic points
$ \mathcal{ D}= \sum_i^n m_i (y_i, x_i) $,  and $ \mathrm{deg} \mathcal{
D}= \sum_i^n m_i$. The { \em effective divisor} is such that $m_i>0
 \forall i$.

Let $ \mathcal{D}$ be a divisor of degree $0$,  $ \mathcal{
D}= \mathcal{ X}- \mathcal{ Z}$,  with $ \mathcal{ X}$ and
$ \mathcal{ Z}$ --- the
effective divisors $ \mathrm{ deg } \; \mathcal{ X}= \mathrm{ deg } \;
 \mathcal{Z}=n$ presented by $ \mathcal{ X}= \{ (y_1,  x_1) ,
\ldots,   (y_n,  x_n)  \}$ and $ \mathcal{ Z}= \{ (w_1,  z_1) ,
\ldots,  (w_n,  z_n)  \}  \in  (V) ^{n}$,  where $ (V) ^n$ is the
$n$--th symmetric power of $V$.

The {  \em Abel map} $  \mathfrak{ A}: (V) ^n   \rightarrow   \mathrm{Jac}
(V) $ puts into correspondence the divisor $  \mathcal{ D}$,  with
fixed $  \mathcal{ Z}$,  and the point $  \boldsymbol{ u}= (u_1  \ldots,
u_g) ^T  \in   \mathrm{Jac} (V) $ according to the   \[  \boldsymbol{
u}=  \int_{  \mathcal{ Z}}^  \mathcal{ X}   \mathrm{d}  \mathbf{ u},
  \quad  \text{or}  \quad u_i=  \sum_{k=1}^n  \int_{z_k}^{x_k}  \mathrm{d}u_i,
  \quad i=1,   \ldots, g.  \]

The {  \em Abel's theorem} says that the points of the
divisors $  \mathcal{ Z}$ and $  \mathcal{ X}$ are respectively the poles and
zeros of a meromorphic function on $V (y, x) $ iff
$  \int^  \mathcal{ X}_  \mathcal{ Z}  \mathrm{d}  \mathbf{ u}=0  \mod
 \Gamma$.
The {  \em Jacobi inversion problem} is
formulated as the problem of inversion of the map $  \mathfrak{ A}$,
when $n=g$ the $  \mathfrak{ A}$ is $1  \to1$,  except for so called
{  \em special divisors}. In our case special divisors  of degree
$g$ are such that  at least for one pair $j$ and $k   \in 1  \ldots
g$ the point $ (y_j, x_j) $ is the image of the hyperelliptic
involution of the  point $ (y_k, x_k) $.

  \subsubsection{Meromorphic differentials} or the differentials of the
second kind,  are the differential $1$-forms $  \mathrm{d}r$ which can
be locally given as
$  \mathrm{d}r= (  \sum_{i=-k}^  \infty  \alpha_i  \xi^i) d  \xi$ in the
vicinity
of any point $ (y, x) $ with some constants $  \alpha_i$,  and
$  \alpha_{ (-1) }=0$.  It can be also checked directly,  that forms
satisfying such a condition are all of the form $
  \sum_{i=0}^{g-1}   \beta_i x^{i+g}  \frac{  \mathrm{d}x}{y}$  ($  \mod$
holomorphic differential) .
Let us introduce the following {  \em canonical Abelian
differentials of the second kind}   \begin{equation}
\mathrm{d}r_j=  \sum_{k=j}^{2g+1-j} (k+1-j)   \lambda_{k+1+j}\frac{x^k
  \mathrm{d}x}{
4y},   \quad j=1,   \ldots, g.   \end{equation} We denote their
 matrices of $A$ and $B$--periods,
  \[2  \eta=  \left (-  \oint_{A_k}  \mathrm{d}r_l  \right) ,   \quad 2  \eta'
=  \left (-  \oint_{B_k}  \mathrm{d}r_l  \right) .  \]
 From {  \em Riemann bilinear
identity},  for the period matrices of the differentials of the
first and second kind follows:    \begin{lemma} $2g  \times
2g$--matrix $  \mathcal{ G}=  \left (  \begin{array} {cc}  \omega&
\omega'  \\
  \eta&  \eta'  \end{array}  \right) $ belongs to $PSp_{2g}$:
  \[  \mathcal{ G}  \left (  \begin{array}{cc}0&-  \mathbf{
1}_g  \\  \mathbf{ 1}_g&0   \end{array}  \right)    \mathcal{ G}^T=-
 \frac{  \pi
i}{2}  \left (  \begin{array}{cc} 0&-  \mathbf{ 1}_g  \\  \mathbf{
1}_g&0  \end{array}  \right) .   \]    \label{LPSR}
  \end{lemma}

  \subsubsection{Differentials of the third kind}
are the differential 1-forms $  \mathrm{d}  \Omega$ to have only
poles of order $1$ and $0$ total residue,  and so are locally
given in the vicinity of any of the poles as
$  \mathrm{d}  \Omega= (  \sum_{i=-1}^  \infty  \alpha_i  \xi^i) d  \xi$
with some constants $  \alpha_i$,  $  \alpha_{-1}$ being nonzero. Such
forms  ($  \mod$  holomorphic differential)  may be presented as:
  \[
  \sum_{i=0}^{n}   \beta_i
  \left (  \frac{y+y^+_i}{x-x^+_i}-  \frac{y+y^-_i}{x-x^-_i}  \right)
  \frac{  \mathrm{d}x}{y},
  \]
where $ (y^  \pm_i, x^  \pm_i) $ are the analytic points of the poles of
positive  (respectively,  negative)  residue.

Let us introduce the canonical differential of the third kind
  \begin{equation}
  \mathrm{d}   \Omega (x_1, x_2) =
  \left (  \frac{y+y_1}{x-x_1}-  \frac{y+y_2}{x-x_2}  \right)
  \frac{  \mathrm{d} x}{2y},   \label{third}
  \end{equation}
for this differential we have $  \int_{x_3}^{x_4}  \mathrm{d}
  \Omega (x_1, x_2) =  \int_{x_1}^{x_2}  \mathrm{d}   \Omega (x_3, x_4) $.

  \subsubsection{Fundamental $2$--differential of the second kind}
For
$  \{ (y_1, x_1) ,  (y_2, x_2)   \}  \in  (V) ^2$
we introduce function  $F (x_1, x_2) $ defined by the conditions
  \begin{eqnarray}
 (  \text{i}) .  &&F (x_1, x_2) =F (x_2, x_1) ,    \nonumber   \\
 (  \text{ii}) .  &&F (x_1, x_1) =2f (x_1) ,   \nonumber   \\
 (  \text{iii}) .  &&  \frac{  \partial
F (x_1, x_2) }{   \partial x_2}   \big|_{x_2=x_1}
=  \frac{  \mathrm{d}f (x_1) }{   \mathrm{d}x_1}.  \label{propf}
  \end{eqnarray}

Such $F (x_1, x_2) $ can be presented in the following
equivalent forms
  \begin{eqnarray}
F (x_1, x_2) &=&2y_2^2+2 (x_1-x_2) y_2  \frac{  \mathrm{d} y_2}{
  \mathrm{d}x_2}  \nonumber  \\
&+& (x_1-x_2) ^2  \sum_{j=1}^gx_1^{j-1}  \sum_{k=j}^{2g+1-j} (k-j+1)
   \lambda_{k+j+1}x_2^k,
  \label{Formf-1}  \\
F (x_1, x_2) &=&2  \lambda_{2g+2}x_1^{g+1}x_2^{g+1}+
  \sum_{i=0}^{g}x_1^ix_2^i (2  \lambda_{2i}+  \lambda_{2i+1} (x_1+x_2) ) .
  \label{Formf-2}   \end{eqnarray}

Properties   \eqref{propf} of  $F (x_1, x_2) $ permit to construct
the {  \em global Abelian $2$--differential of the second
kind} with the unique  pole of order $2$  along $x_1=x_2$ :
  \begin{equation}
  \omega (x_1, x_2) =  \frac{2y_1 y_2+F (x_1, x_2) }{
4 (x_1-x_2) ^2}  \frac{  \mathrm{d}x_1}{ y_1}  \frac{  \mathrm{d}x_2}{y_2},
  \label{omega-1}   \end{equation} which expands in the vicinity of
the pole as   \[
  \omega (x_1, x_2) =  \left (  \frac{1}{2 (  \xi-  \zeta) ^2}+O (1)
  \right)   \mathrm{d}  \xi
  \mathrm{d}  \zeta,
  \]
where $  \xi$ and $  \zeta$ are the local
coordinates at the points $x_1$ and $x_2$ correspondingly.

Using the   \eqref{Formf-1},  rewrite the   \eqref{omega-1} in the form
  \begin{equation}
  \omega (x_1, x_2) =  \frac{  \partial}{   \partial x_2}
\left (  \frac{y_1+y_2}{
2y_1 (x_1-x_2) }   \right)   \mathrm{d}x_1  \mathrm{d}x_2
+  \mathrm{d}  \mathbf{
u}^T (x_1)   \mathrm{d}  \mathbf{ r} (x_2) ,   \label{omega-2}  \end{equation}
where the differentials $  \mathrm{d}  \mathbf{ u},   \mathrm{d}  \mathbf{ r}$
are as above. So,  the  periods of this $2$-form  (the double
integrals $  \oint  \oint  \omega (x_1, x_2) $)  are expressible in terms of
$ (2  \omega, 2  \omega') $ and $ (-2  \eta, -2  \eta') $,  e.g.,  we
have for $A$-periods:
  \[
  \left  \{  \oint_{A_i}  \oint_{A_k}  \omega (x_1, x_2)   \right  \}_{i, k=1,   \ldots, g}=
-4  \omega^T  \eta.
  \]

  \subsection{Riemann $  \theta$-function}

The standard $  \theta$--function $  \theta (  \boldsymbol{ v}|  \tau) $ on $  \mathbb{
C}^g  \times   \mathcal{ S}_g$ is defined by its Fourier series,

  \[
  \theta (  \boldsymbol{ v}|  \tau)
=  \sum_{  \boldsymbol{ m}  \in   \mathbb{ Z}^g}  \mathrm{
exp}  \;  \pi i  \left  \{   \boldsymbol{ m}^T  \tau   \boldsymbol{ m}
+2  \boldsymbol{ v}^T
  \boldsymbol{ m}  \right  \}  \]
The $  \theta$--function
possesses the periodicity properties $  \forall k   \in 1,   \ldots, g$
  \begin{eqnarray*} &&  \theta (v_1,   \ldots, v_k+1,   \ldots, v_g|  \tau)
=  \theta (  \boldsymbol{ v}|  \tau) ,   \\
&&  \theta (v_1+  \tau_{1k},   \ldots, v_k+  \tau_{kk},   \ldots, v_g
+  \tau_{gk}|  \tau)
=  \mathrm{ e}^{i  \pi   \tau_{kk}-2  \pi i v_k}  \theta (  \boldsymbol{ v}|  \tau) .
  \end{eqnarray*}

$  \theta$--functions with characteristics $[  \varepsilon]=
  \left[   \begin{array}{c}  \varepsilon'  \\
\varepsilon   \end{array}  \right]
=  \left[   \begin{array}{ccc}  \varepsilon_1'&
\ldots&  \varepsilon_g'
  \\  \varepsilon_1&  \ldots&  \varepsilon_g
\end{array}  \right]   \in   \mathbb{
C}^{2g}$
  \[
  \theta[  \varepsilon] (  \boldsymbol{ v}|  \tau) =  \sum_{  \boldsymbol{ m}  \in   \mathbb{
Z}^g}  \mathrm{ exp}  \;  \pi i  \left  \{  (  \boldsymbol{ m}+  \varepsilon') ^T  \tau
 (  \boldsymbol{ m}+  \varepsilon') +2 (  \boldsymbol{ v}+  \varepsilon) ^T
 (  \boldsymbol{ m}+  \varepsilon')   \right  \},
  \] for which
the periodicity properties are
  \begin{eqnarray*}
&&  \theta[  \varepsilon] (v_1,   \ldots, v_k+1,   \ldots, v_g|  \tau)
=  \mathrm{ e}^{2  \pi i  \varepsilon_k'}  \theta (  \boldsymbol{
v}|  \tau) ,   \\
&&  \theta[  \varepsilon] (v_1+  \tau_{1k},   \ldots, v_k+  \tau_{kk},\ldots, v_g
+  \tau_{gk}|  \tau)
=  \mathrm{ e}^{i  \pi   \tau_{kk}-2  \pi i v_k-2  \pi i
  \varepsilon_k}  \theta (  \boldsymbol{ v}|  \tau) .
  \end{eqnarray*}

Further,  consider half-integer characteristics $[  \varepsilon]$;
the $  \theta$--function $  \theta[  \varepsilon] (  \boldsymbol{ v}|\tau) $
is even or odd whenever $4{  \varepsilon'}^T  \varepsilon=0$ or 1
modulo 2.  There are $  \frac{1}{2} (4^g+2^g) $ even characteristics
and $  \frac{1}{2} (4^g-2^g) $ odd.

Let $  \boldsymbol{ w}^T= (w_1,   \ldots, w_g)   \in   \mathrm{Jac} (V) $ be
some fixed vector,  the function,
  \[   \mathcal{
 R} (x) =  \theta  \left (   \int_{x_0}^x   \mathrm{d}  \mathbf{ v} -  \boldsymbol{
w}|  \tau  \right) ,   \quad x  \in V
  \] is called {  \em Riemann $  \theta$--function}.

The Riemann $  \theta$--function $  \mathcal{ R} (x) $ is either identically
$0$,  or it has exactly $g$ zeros $x_1,   \ldots, x_g  \in V$,  for which
the {  \em Riemann vanishing theorem} says that
  \[
  \sum_{k=1}^g
  \int_{x_0}^{x_i}   \mathrm{d}  \mathbf{ v}=  \boldsymbol{ w}+  \mathbf{
K}_{x_0},   \] where $  \mathbf{ K}^T_{x_0}= (K_1,   \ldots, K_g) $ is the
vector of Riemann constants with  respect to the base point
$x_0$ and is defined by the formula
  \begin{equation}
K_j=  \frac{1+  \tau_{jj}}{ 2}-
  \sum_{l  \neq j}
  \oint_{A_l} \mathrm{d}v_l (x)   \int_{x_0}^x \mathrm{d}v_j,
  \quad j=1,   \ldots,  g.
  \label{Rconst}   \end{equation}

  \section{Kleinian functions}  \label{sigma-defs}
Let $  \boldsymbol{ m},   \boldsymbol{ m}'  \in   \mathbb{ Z}^g$ be two
arbitrary vectors; denote  periods $  \mathbf{ E} (  \boldsymbol{ m},
  \boldsymbol{
m}') =2  \eta   \boldsymbol{ m}+2  \eta'  \boldsymbol{ m}'$,  $  \boldsymbol{   \Omega} (
  \boldsymbol{
m},   \boldsymbol{ m}') =2  \omega   \boldsymbol{ m}+2  \omega'  \boldsymbol{ m}'$.
  \subsection{$  \sigma$--function}

In    \cite{kl88, ba98}  it  was shown,  that the properties
  \eqref{period} and   \eqref{0-expan}
 define the function,  which plays the central role in the
theory of Kleinian functions.

  \begin{definition} An integral function
$  \sigma (  \boldsymbol{ u}) $ is the Kleinian {  \em fundamental
$  \sigma$--function} iff
  \begin{enumerate}   \item for any vector
$  \boldsymbol{ u}  \in   \mathrm{Jac} (V) $   \begin{equation}   \sigma (
  \boldsymbol{
u}+  \boldsymbol{   \Omega} (  \boldsymbol{  m},   \boldsymbol{  m}') )
=  \mathrm{
exp}  \left  \{   \mathbf{ E}^T (  \boldsymbol{  m},   \boldsymbol{  m}')  (  \boldsymbol{
u}+  \tfrac{1}{2}  \boldsymbol{   \Omega} (  \boldsymbol{
m},   \boldsymbol{ m}') )  +  \pi i  \boldsymbol{  m}^T   \boldsymbol{
m}'  \right  \}  \sigma (  \boldsymbol{  u}) .    \label{period}
  \end{equation}
  \item $  \sigma (  \boldsymbol{  u}) $ has $0$ of
$  \left[  \frac{g+1}{2}  \right]$ order at $  \boldsymbol{  u}=0$ and
  \begin{equation}   \lim_{  \boldsymbol{  u}  \to 0}  \frac{  \sigma (  \boldsymbol{
u}) }{  \delta (  \boldsymbol{  u}) }=1,   \label{0-expan}   \end{equation} where
$  \delta (  \boldsymbol{  u}) =
  \det  \left (-  \{u_{i+j-1}  \}_{i, j=1,   \ldots,   \left[  \frac{g+1}{2}  \right]}  \right) $.
  \end{enumerate}
  \end{definition}

For small genera we have,
$  \sigma=u_1+  \ldots$ for $g=1$ and $2$;
$  \sigma=u_1u_3-u_2^2+  \ldots$ for $g=3$ and $4$;
$  \sigma=-u_3^3+2u_2u_3u_4-u_1u_4^2-u_2^2u_5+u_1u_3u_5+  \ldots$ for $g=5$
and $6$  etc.

We introduce the $  \sigma$-{  \em functions with characteristic},
$  \sigma_{  \boldsymbol{  r},   \boldsymbol{  r}'}$ for vectors
 $   \boldsymbol{  r}^T,    \boldsymbol{  r}'  \in   \frac12   \mathbb{ Z}^g/  \mathbb{ Z}^g$ defined
by the formula
  \[
  \sigma_{  \boldsymbol{  r},   \boldsymbol{  r}'} (  \boldsymbol{  u}) =  \mathrm{
e}^{-  \mathbf{E}^T (  \boldsymbol{  r},   \boldsymbol{  r}')   \boldsymbol{
u}}  \frac{  \sigma (  \boldsymbol{ u}+  \boldsymbol{   \Omega} (  \boldsymbol{
r},   \boldsymbol{  r}') ) } {  \sigma (  \boldsymbol{   \Omega} (  \boldsymbol{
r},   \boldsymbol{ r}') ) }.   \] These functions are completely
analogous to the Weierstrass' $  \sigma_  \alpha$  appearing in the
elliptic theory   \cite{ba55}.

  \subsubsection{ $  \sigma$--function as
 $  \theta$--function}
Fundamental hyperelliptic Kleinian $  \sigma$--function belongs to
the class of generalized $  \theta$--functions.  We give the
explicit expression of the $  \sigma $ in terms of standard
$  \theta$--function as follows:
\begin{equation}   \sigma (  \boldsymbol{
u}) =C  \mathrm{ e}^{  \boldsymbol{  u}^T   \varkappa    \boldsymbol{
u}}
\theta ( (2  \omega) ^{-1}
\boldsymbol{u}-
\mathbf{K}_{a}|\tau) ,
\label{sigma}
\end{equation}
where $   \varkappa = (2  \omega) ^{-1}  \eta$,
$  \mathbf{ K}_{a}$ is the vector of Riemann constants with the base
point $a$ and the constant    \[
C=  \frac{  \epsilon_4}{  \theta (0|  \tau) }
  \prod_{r=1}^g  \frac{  \sqrt{{P}' (a_r) }}{  \sqrt[4]{f' (a_r) }}
  \frac{1}{  \prod_{k<l}  \sqrt{e_k-e_l}},   \]
where $ (  \epsilon_4) ^4=1$.

Direct calculation shows,  that the function defined by
  \eqref{sigma} satisfies   \eqref{period} and
  \eqref{0-expan}, we only note that,
in  our case the vector of Riemann constants
  \eqref{Rconst} is, as follows from Riemann vanishing theorem,
  \begin{equation}
  \mathbf{ K}_{a}=  \sum_{k=1}^g  \int_{a}^{a_i}
  \mathrm{d}  \mathbf{ v}.   \label{rvector}
\end{equation}

Putting $g=1$ and fixing the elliptic curve $y^2=f (x) =4x^3-g_2
x-g_3$ in   \eqref{sigma},  we see that the function
  \[  \sigma (u) =  \frac{1}{   \vartheta_3 (0|  \tau)
  \sqrt[4]{ (e_1-e_2)  (e_2-e_3) }}   \mathrm{ e}^  \frac{  \eta u^2}{
2  \omega}  \vartheta_1   \left (  \frac{u}{ 2  \omega}  \big|  \tau  \right)   \]
 is the standard Weierstrass $  \sigma$--function,  were we have used
the standard notation for Jacobi $  \vartheta$--functions  (see e.g.
  \cite{ba55} ) .

  \subsection{Functions $  \zeta$ and $  \wp$}
Kleinian $  \zeta$ and $  \wp$-functions  are defined as logarithmic
derivatives of the fundamental $  \sigma$
  \begin{eqnarray*}&&  \zeta_i (  \boldsymbol{  u}) =  \frac{  \partial
\mathrm{
ln}  \;  \sigma (  \boldsymbol{  u})  }{   \partial u_i},   \quad
i  \in 1,   \ldots, g;  \\ &&  \wp_{ij} (  \boldsymbol{  u}) =-  \frac{  \partial^2
  \mathrm{ ln}  \;   \sigma (  \boldsymbol{ u}) }{   \partial u_i  \partial
u_j},   \;   \wp_{ijk} (  \boldsymbol{ u}) =-  \frac{  \partial^3   \mathrm{
ln}  \;  \sigma (  \boldsymbol{  u}) }{   \partial u_i  \partial
 u_i  \partial  u_k}  \ldots,  i, j, k,   \ldots  \in 1,   \ldots, g.
  \end{eqnarray*}

The functions $  \zeta_i (  \boldsymbol{  u}) $ and $  \wp_{ij} (  \boldsymbol{  u}) $ have the
following periodicity properties
  \begin{eqnarray*}   \zeta_i (  \boldsymbol{
u}+  \boldsymbol{   \Omega} (  \boldsymbol{  m},   \boldsymbol{  m}') ) &=
&  \zeta_i (  \boldsymbol{  u})
+E_i (  \boldsymbol{  m},   \boldsymbol{  m}') ,    \quad i  \in 1,   \ldots, g,
  \\   \wp_{ij} (  \boldsymbol{  u}+
  \boldsymbol{   \Omega} (  \boldsymbol{  m},   \boldsymbol{  m}') )
&=&  \wp_{ij} (  \boldsymbol{  u}) ,    \quad
i, j   \in 1,   \ldots, g.   \end{eqnarray*}

\subsubsection{Realization of the fundamental $2$--differential of
the second kind by Kleinian functions}
The construction is based on the following
  \begin{theorem}
  \label{the-S} Let $ (y (a_0),  a_0) $,  $ (y, x) $ and $ (  \nu,
\mu) $  be arbitrary distinct  points on $V$ and let $  \{ (y_1,
x_1),   \ldots,  (y_g, x_g)   \}$  and $  \{ (  \nu_1, \mu_1) ,
 \ldots, (  \nu_g,   \mu_g)   \}$ be arbitrary sets of distinct
 points  $  \in (V) ^g$.  Then the following relation is valid
 \begin{eqnarray} \lefteqn{   \int_{  \mu}^x \sum_{i=1}^g   \int_{
  \mu_i}^{x_i} \frac{2yy_i+F (x, x_i) }{4 (x-x_i) ^2}   \frac{
  \mathrm{d}x}{y} \frac{ \mathrm{d}x_i}{y_i}}  \nonumber  \\ &&=
  \mathrm{ln }   \; \left  \{ \frac{  \sigma   \left (
\int_{a_0}^x  \mathrm{d} \mathbf{ u} -  \sum_{i=1}^g
  \int_{a_i}^{x_i}  \mathrm{d} \mathbf{ u}  \right)  }{  \sigma
\left (  \int_{a_0}^x  \mathrm{d} \mathbf{ u}-  \sum_{i=1}^g
  \int_{a_i}^{  \mu_i}  \mathrm{d} \mathbf{ u}  \right) } \right
\}-  \mathrm{ ln }   \;   \left  \{ \frac{  \sigma \left (
  \int_{a_0}^{  \mu}  \mathrm{d}  \mathbf{ u}-  \sum_{i=1}^g
\int_{a_i}^{x_{i}}  \mathrm{d}  \mathbf{ u}  \right) }{  \sigma
  \left (  \int_{a_0}^{  \mu}  \mathrm{d}  \mathbf{
u}-  \sum_{i=1}^g  \int_{a_i}^{  \mu_i}  \mathrm{d}  \mathbf{
u}  \right) }  \right  \} ,   \label{r11}  \end{eqnarray} where the function
$F (x, z) $ is given by   \eqref{Formf-2}.    \end{theorem}

  \begin{proof} Let us consider the sum
  \begin{equation}
  \sum_{i=1}^g  \int_{  \mu}^x  \int_{  \mu_i}^{x_i}   \left[
  \omega (x, x_i)  +
  \mathrm{d}  \mathbf{ u}^T (x)
  \varkappa
  \mathrm{d}  \mathbf{ u} (x_i)   \right],
  \label{r10}
  \end{equation}
with $  \omega (  \cdot,   \cdot) $ given by   \eqref{omega-2}.
It is the normalized Abelian
integral of the third kind with the logarithmic residues in the
points $x_i$ and $  \mu_i$. By Riemann vanishing theorem  we can
express   \eqref{r10} in terms of Riemann $  \theta$--functions as
  \begin{equation}
  \mathrm{ln}
  \left  \{
  \tfrac{{  \displaystyle{  \theta}}   \left (  \int_{a_0}^x   \mathrm{d}  \mathbf{
v}- (  \sum_{i=1}^g  \int_{a_0}^{x_i}  \mathrm{d}  \mathbf{ v}-  \mathbf{
K}_{a_0})   \right) }
{{  \displaystyle{  \theta}}   \left (  \int_{a_0}^{x}  \mathrm{d}  \mathbf{
v}- (  \sum_{i=1}^g  \int_{a_0}^{  \mu_i}  \mathrm{d}  \mathbf{ v}-  \mathbf{
K}_{a_0})   \right) }
  \right  \}
-  \mathrm{ln}
  \left  \{
  \tfrac{{  \displaystyle{  \theta}}
  \left (  \int_{a_0}^{  \mu}  \mathrm{d}  \mathbf{
 v}- (  \sum_{i=1}^g  \int_{a_0}^{x_i}  \mathrm{d}  \mathbf{ v}-  \mathbf{
K}_{a_0})   \right) }{{  \displaystyle{  \theta}}
  \left (  \int_{a_0}^{  \mu}  \mathrm{d}  \mathbf{
v}- (  \sum_{i=1}^g  \int_{a_0}^{  \mu_i}  \mathrm{d}  \mathbf{ v}-  \mathbf{
K}_{a_0})   \right) }  \right  \},   \label{in-depends}
  \end{equation}
and to obtain right hand side of   \eqref{r11} we have to combine the
  \eqref{sigma},   expression of the vector $  \mathbf{ K}_{a_0}$
  \eqref{rvector},  matrix $   \varkappa = (2  \omega) ^{-1}  \eta$
and Lemma   \ref{LPSR}. Left hand side  of
  \eqref{r11} is obtained using   \eqref{omega-1}.   \end{proof}

 The fact,  that right hand side of the
  \eqref{r11} is independent on the arbitrary point $a_0$,  to be
 employed further,  has its origin in the properties of the vector
 of Riemann constants. Consider the difference $  \mathbf{ K}_{a_0}-
   \mathbf{ K}_{a_0'}$ of vectors of Riemann constants with arbitrary
 base points $a_0$ and $a_0'$ by   \eqref{Rconst} we find
   \[
   \mathbf{ K}_{a_0}-
   \mathbf{ K}_{a_0'} = (g-1)   \int_{a_0'}^{a_0}  \mathrm{d}  \mathbf{ v},
   \]
this property provides that
   \[
  \int_{a_0}^{x_0}  \mathrm{d}  \mathbf{
 v}- (  \sum_{i=1}^g  \int_{a_0}^{x_i}  \mathrm{d}  \mathbf{ v}-  \mathbf{
K}_{a_0}) =
   \int_{a_0'}^{x_0}  \mathrm{d}  \mathbf{
 v}- (  \sum_{i=1}^g  \int_{a_0'}^{x_i}  \mathrm{d}  \mathbf{ v}-  \mathbf{
K}_{a_0'})
   \]
for arbitrary $x_i, $ with $i  \in 0,   \ldots, g$  on $V$,  so the
arguments of $  \sigma$'s in   \eqref{r11} which are linear
transformations by $2  \omega$ of the arguments of $  \theta$'s in
\eqref{in-depends},  do not depend on $a_0$.

  \begin{cor}  \label{cor-P}  From Theorem   \ref{the-S} for arbitrary
distinct $ (y (a_0) , a_0) $ and $ (y, x) $ on $V$ and  arbitrary set of
distinct points $  \{ (y_1, x_1)   \ldots,  (y_g, x_g)   \}  \in  (V) ^g$ follows:
  \begin{equation}   \sum_{i, j=1}^g  \wp_{ij}   \left (  \int_{a_0}^x
  \mathrm{d}  \mathbf{ u}+   \sum_{k=1}^g  \int_{a_k}^{x_k}  \mathrm{d}  \mathbf{
u}  \right)  x^{i-1}x_r^{j-1}=  \frac{F (x, x_r) -2yy_r}{ 4 (x-x_r) ^2},   \;
r=1,   \ldots, g.  \label{principal}
  \end{equation}
  \end{cor}
  \begin{proof}
Taking the partial derivative $  \partial^2/  \partial x_r  \partial x$
from the both sides of   \eqref{r11} and using the hyperelliptic
involution $  \phi (y, x) = (-y, x) $ and
$  \phi (y (a_0) , a_0) = (-y (a_0) , a_0) ) $ we obtain   \eqref{principal}.  \end{proof}

In the case $g=1$ the formula   \eqref{principal} is
actually the addition theorem for the Weierstrass elliptic
functions,
  \[
  \wp (u+v) =-  \wp (u) -  \wp (v) +  \frac{1}{4}  \left[  \frac{  \wp' (u) -  \wp' (v) }{
  \wp (u) -  \wp (v)  }  \right]^2
  \]
on the  elliptic curve  $y^2=f (x) =4x^3-g_2
x-g_3$.

Now we can give the expression for $  \omega (x, x_r) $ in terms of
Kleinian functions.  We send the base point $a_0$ to the branch
 place $a$,  and  for $r  \in 1,   \ldots, g $ the fundamental
$2$--differential of the second kind is given by   \[   \omega (x, x_r) =
  \sum_{i, j=1}^g  \wp_{ij}   \left (  \int_a^x   \mathrm{d}  \mathbf{ u}-
  \sum_{k=1}^g  \int_{a_k}^{x_k}  \mathrm{d}  \mathbf{
u}  \right)   \frac{x^{i-1}  \mathrm{d}x}{y}  \frac{x_r^{j-1}
\mathrm{d}x_r}{y_r}. \]
  \begin{cor}
  $  \forall r   \neq
s   \in 1,   \ldots, g$
  \begin{eqnarray}
  \sum_{i, j=1}^g  \wp_{ij}
  \left (  \sum_{k=1}^g  \int_{a_k}^{x_k}  \mathrm{d}  \mathbf{ u}  \right)
x_s^{i-1}x_r^{j-1} =  \frac{F (x_s, x_r) -2y_sy_r}{ 4 (x_s-x_r) ^2}
.  \label{principal11}
  \end{eqnarray}
  \end{cor}
  \begin{proof}
In   \eqref{principal} we have for $s  \neq r$
  \begin{eqnarray*}
&  \int  \limits_{  \phi (a_0) }^x
  \mathrm{d}  \mathbf{ u}+   \sum  \limits_{k=1}^g
  \int  \limits_{a_k}^{x_k}  \mathrm{d}  \mathbf{
u}
 =-2  \omega (  \int  \limits_{a_0}^{  \phi (x) }  \mathrm{d}  \mathbf{
 v}- (  \sum  \limits_{i=1}^g  \int  \limits_{a_0}^{x_i}  \mathrm{d}
\mathbf{ v}-  \mathbf{
K}_{a_0}) ) =  \\
&-2  \omega (  \int  \limits_{x_s}^{  \phi (x) }  \mathrm{d}  \mathbf{
 v}- (  \sum  \limits_{  \begin{subarray}{l}i=1  \\
i  \neq s  \end{subarray}}^g  \int  \limits_{x_s}^{x_i}  \mathrm{d}  \mathbf{
v}-  \mathbf{ K}_{x_s}) ) =  \int  \limits_{a_s}^{x}  \mathrm{d}  \mathbf{
 u}+  \sum  \limits_{  \begin{subarray}{l}i=1  \\
i  \neq s  \end{subarray}}^g  \int  \limits_{a_i}^{x_i}  \mathrm{d}  \mathbf{
u}
  \end{eqnarray*}
and the change of notation $x   \to x_s$ gives
  \eqref{principal11}.    \end{proof}

  \subsubsection{Solution of the Jacobi inversion problem}
The equations of Abel map in conditions of Jacobi
inversion problem
  \begin{equation}
u_i=  \sum_{k=1}^g  \int_{a_k}^{x_k}   \frac{x^{i-1}  \mathrm{d}x}{y},
  \label{Abel-map}
  \end{equation}
are invertible if the points $ (y_k,  x_k) $ are distinct and
$  \forall j, k   \in 1,   \ldots, g  \;   \phi (y_k, x_k)   \neq (y_j, x_j)  $.  Using
  \eqref{principal} we find the solution of Jacobi inversion problem
on the curves with $a=  \infty$ in a very effective form.

  \begin{theorem} The Abel preimage of the point $  \boldsymbol{
u}  \in  \mathrm{Jac} (V) $ is given  by the set
$  \{ (y_1, x_1) ,   \ldots,  (y_g, x_g)   \}  \in  (V) ^g$,  where
$  \{x_1,   \ldots, x_g  \}$ are the zeros of the polynomial
  \begin{equation}   \mathcal{ P} (x;  \boldsymbol{  u}) =0  \label{x},
 \end{equation}
where   \begin{equation}   \mathcal{ P} (x;  \boldsymbol{  u}) =x^g-x^{g-1}
 \wp_{g, g}
 (  \boldsymbol{ u}) -x^{g-2}  \wp_{g, g-1} (  \boldsymbol{ u}) -  \ldots-
\wp_{g, 1} (  \mathbf{
u}) ,   \label{p}
  \end{equation}
and  $  \{y_1,   \ldots, y_g  \}$ are given by
  \begin{equation} y_k=-  \frac{  \partial   \mathcal{
P} (x;  \boldsymbol{  u}) }{   \partial u_g}  \Bigl\lvert_{x=x_k},
  \; \label{y} \end{equation} \end{theorem} \begin{proof} We tend
in \eqref{principal} $a_0  \to a=  \infty$. Then we take
  \begin{equation}
  \lim_{x   \to  \infty}
  \frac{F (x, x_r) }{4x^{g-1} (x-x_r) ^2}=
  \sum_{i=1}^g  \wp_{gi} (  \boldsymbol{ u}) x_r^{i-1}.  \label{limit}
  \end{equation} The limit in the left hand side of  \eqref{limit} is
equal to $x_r^g$,  and we obtain   \eqref{x}.

We find from   \eqref{Abel-map},
  \[  \sum_{i=1}^g  \frac{x^{k-1}_i}{ y_i}
  \frac{  \partial x_i}{   \partial u_j}=  \delta_{jk},   \qquad
\frac{  \partial x_k}{   \partial u_g}=  \frac{y_k
}{  \prod_{i  \neq
k} (x_k-x_i) }.  \]
On the other hand we have
  \[
  \frac{  \partial   \mathcal{ P}}{   \partial u_g}
\Bigl\lvert_{x=x_k}= -  \frac{  \partial x_k}{   \partial u_g}
\prod_{i  \neq k} (x_i-x_k) ,   \] and we obtain   \eqref{y}.
\end{proof}

Let us denote by $  \boldsymbol{  \wp}$,  $  \boldsymbol{  \wp}'$ the
$g$--dimensional vectors,
  \[
  \boldsymbol{   \wp}=  \left (  \wp_{g1},   \ldots,   \wp_{gg}
  \right) ^T,
  \quad
  \boldsymbol{   \wp}'=  \frac{  \partial   \boldsymbol{   \wp}}{
\partial u_g}
  \] and the companion matrix   \cite{hj86} of the polynomial
$  \mathcal{ P} (z;  \boldsymbol{ u}) $,  given by   \eqref{p}
  \[  \mathcal{ C}=  \mathcal{ B}+   \boldsymbol{   \wp}   \mathbf{ e}_g^T,
  \quad  \text{where}  \quad  \mathcal{ B}=   \sum_{k=1}^g   \mathbf{ e}_k
 \mathbf{
e}^T_{k-1}.  \]   The companion matrix $  \mathcal{
C}$ has the property   \begin{equation} x_k^n=  \mathbf{ X}^T_k  \mathcal{
C}^{n-g+1}  \mathbf{ e}_g=  \mathbf{ X}^T_k  \mathcal{ C}^{n-g}  \boldsymbol{
  \wp},    \quad   \forall n  \in   \mathbb{ Z},    \label{prop-c}
\end{equation} with the vector $  \mathbf{ X}_k^T= (1, x_k,
\ldots, x_k^{g-1}) $,  where $x_k$ is one of the roots of
\eqref{x}.  From   \eqref{principal11} we find $-2y_ry_s=4
(x_r-x_s) ^2  \sum_{i=1}^{g}  \wp_{ij} (  \boldsymbol{ u})
x_r^{i-1}x_s^{j-1}-F (x_r, x_s) $.  Introducing matrices $  \Pi= (
\wp_{ij}) $, $  \Lambda_0=  \mathrm{diag} (  \lambda_{2g-2},
\ldots,   \lambda_0) $ and $  \Lambda_1=  \mathrm{diag}  \; (
\lambda_{2g-1},   \ldots,   \lambda_1) $, we have,  taking into
 account   \eqref{prop-c}, \begin{eqnarray*} \lefteqn{2  \mathbf{
  X}^T_r  \boldsymbol{ \wp}'{  \boldsymbol{   \wp}'}^T  \mathbf{
  X}_s=-4  \mathbf{ X}^T_r ({  \mathcal{ C}}^2  \Pi-2   \mathcal{
C}  \Pi   \mathcal{ C}^T+  \Pi {  \mathcal{ C}^T}^2 )    \mathbf{
X}_s }   \nonumber  \\&&+ 4  \mathbf{ X}^T_r (  \mathcal{ C}
\boldsymbol{ \wp}  \boldsymbol{   \wp}^T+  \boldsymbol{   \wp}
  \boldsymbol{   \wp}^T  \mathcal{ C}^T)   \mathbf{ X}_s +2
\mathbf{ X}^T_r  \Lambda_0  \mathbf{ X}_s+  \mathbf{ X}^T_r (
\mathcal{ C}  \Lambda_1+  \Lambda_1  \mathcal{ C}^T)   \mathbf{
X}_s.  \end{eqnarray*} Whence,   (see   \cite{be96b}) :
  \begin{cor}
The relation
  \begin{equation}
2  \boldsymbol{
  \wp}'{  \boldsymbol{   \wp}'}^T=-4 ({  \mathcal{
C}}^2  \Pi-2   \mathcal{ C}  \Pi   \mathcal{ C}^T+  \Pi {  \mathcal{ C}^T}^2 )
+ 4 (  \mathcal{ C}  \boldsymbol{
  \wp}  \boldsymbol{   \wp}^T+  \boldsymbol{   \wp}  \boldsymbol{   \wp}^T
  \mathcal{
C}^T) +  \mathcal{ C}  \Lambda_1+  \Lambda_1  \mathcal{
C}^T+2  \Lambda_0.  \label{principium}
  \end{equation}
connects odd functions $  \wp_{ggi}$ with poles order $3$ and  even
functions $  \wp_{jk}$ with poles of order $2$ in the field of
meromorphic functions on $  \mathrm{Jac} (V) $.
  \end{cor}

  \begin{definition}  \label{umbral_D} The {  \em umbral derivative}
 \cite{ro84}
  $D_s (p (z) ) $ of a polynomial   \newline $p (z) =  \sum_{k=0}^n p_k z^k$
is given by   \[ D_s p (z) =
  \left (  \frac{p (z) }{z^s}  \right) _+=  \sum_{k=s}^{n}p_kx^{k-s},
  \]
where $ (  \cdot) _+$ means taking the purely polynomial part.
  \end{definition}

Considering polynomials $p=  \prod_{k=1}^{n} (z-z_k) $
and ${  \tilde p}= (z-z_0) p$,  the elementary properties of $D_s$ are
immediately deduced:
  \begin{eqnarray}
&D_s (p) =z D_{s+1} (p) +p_s=z D_{s+1} (p)  +
S_{n-s} (z_1,   \ldots, z_n) ,   \nonumber  \\
&D_s ({  \tilde
p}) = (z-z_0) D_s (p) +p_{s-1}= (z-z_0) D_s (p) +S_{n+1-s} (z_1,
  \ldots, z_n) ,   \label{umbral_2}
  \end{eqnarray}
where $S_l (  \cdots) $ is the $l$--th order elementary symmetric
function of its variables  times $ (-1) ^l$  (we assume
$S_0 (  \cdots) =1$) .

From   \eqref{umbral_2} we see that $S_{n-s} (z_0,   \ldots,   \hat
z_l,   \ldots, z_n) =   \left (D_{s+1} ({  \tilde
p}) |_{z=z_l}  \right) $. This is particularly useful to  write
down the inversion of   \eqref{y}
  \begin{equation}
  \wp_{ggk} (  \boldsymbol{ u}) =  \sum_{l=1}^g y_l
  \left (  \frac{D_k (P (z) ) }
{  \frac{  \partial}{  \partial z}P (z) }  \Bigg|_{z=x_l}  \right) ,
  \label{inversiony}
  \end{equation}
where $P (z) =  \prod_{k=1}^{g} (z-x_k) $.

It is of importance to describe the set of common zeros of the
functions $  \wp_{ggk} (  \boldsymbol{ u}) $.

  \begin{cor}
The vector function $  \boldsymbol{   \wp}' (  \boldsymbol{ u}) $
vanishes iff
$  \boldsymbol{ u}$ is a halfperiod.   \end{cor}

  \begin{proof} The equations $  \wp_{ggk} (  \boldsymbol{ u}) =0,
  \;k  \in 1,   \ldots,  g$ yield due to   \eqref{inversiony} the equalities
$y_i=0,   \forall i  \in 1,   \ldots, g$. The latter is possible if and
only if the points $x_1,   \ldots,  x_g$ coincide with any $g$ points
$e_{i_1},   \ldots, e_{i_g}$ from the set  branching points
$e_1,   \ldots, e_{2g+2}$. So the point
  \[
  \boldsymbol{ u}=  \sum_{l=1}^{g}  \int_{a_l}^{e_{i_l}}  \mathrm{d}  \mathbf{
u}  \in   \mathrm{Jac} (V)    \] is of the second order in Jacobian and
hence is a halfperiod.  \end{proof}

  \section{Basic relations}

In this section  we are going to derive the explicit algebraic
relations between the generating functions in the field of
meromorphic functions on $  \mathrm{Jac} (V) $. After some
preparations just below,  we will in the section~  \ref{hyper-Jac}
find the explicit cubic relations between $  \wp_{ggi}$
and $  \wp_{ij}$.  These,  in turn,  lead
to very special corollaries:  the variety
$  \mathrm{Kum} (V) =  \mathrm{Jac} (V) /  \pm$ is mapped into the space of
symmetric matrices of rank not greater than $3$.

We start with,  the conditions
$  \lambda_{2g+2}=0,   \,   \lambda_{2g+1}=4$ being imposed,
the following Theorem   \ref{the-Z},  which is the starting point for
derivation of the basic relations.

  \begin{theorem}  \label{the-Z}
 Let $(y_0, x_0)\in V$ be an arbitrary point and
$  \{ (y_1, x_1) ,   \ldots,  (y_g, x_g)   \}  \in  (V)^g$ be the
Abel preimage of the point $  \boldsymbol{ u}  \in
\mathrm{Jac} (V) $. Then \begin{equation} -  \zeta_j \left (
\int_{a}^{x_0} \mathrm{d}  \mathbf{ u} +  \boldsymbol{ u}  \right)
  = \int_{a}^{x_0}  \mathrm{d}r_j + \sum_{k=1}^g  \int_{a_k}^{x_k}
  \mathrm{d}r_j -  \frac{1}{2}  \sum_{k=0}^g y_k  \left (
\frac{D_j (R' (z) ) }{R' (z) }  \Bigg|_{z=x_k}  \right) ,
  \label{principalz}
  \end{equation}
where $ R (z) =  \prod_0^g (z-x_j) $ and $
R' (z) =  \frac{  \partial}{  \partial z} R (z) $.

And
  \begin{eqnarray}
-  \zeta_j (  \boldsymbol{ u}) =   \sum_{k=1}^g  \int_{a_k}^{x_k}  \mathrm{d}
r_j-  \frac{1}{2}  \wp_{gg, j+1} (  \boldsymbol{ u}).
  \label{principalz1} \end{eqnarray} \end{theorem}

  \begin{proof} Putting in   \eqref{r11} $  \mu_i=a_i$ we have

  \begin{equation}   \mathrm{ln}
  \left  \{  \frac{  \sigma (  \int_{a_0}^{x}  \mathrm{d}   \mathbf{ u}-  \boldsymbol{ u}) }
{  \sigma (  \int_{a_0}^{x}  \mathrm{d}  \mathbf{ u}) }   \right  \}
-  \left  \{  \frac{  \sigma (  \int_{a_0}^{  \mu}  \mathrm{d}   \mathbf{ u}-  \boldsymbol{ u}) }
{  \sigma (  \int_{a_0}^{  \mu}  \mathrm{d}  \mathbf{ u}) }   \right  \} =
  \int_{  \mu}^{x}  \mathrm{d}  \mathbf{ r}^T  \,   \boldsymbol{
u}+  \sum_{k=1}^g  \int_{a_k}^{x_k}  \mathrm{d}  \Omega (x,   \mu) ,
  \label{r20}  \end{equation}
where $  \mathrm{d}  \Omega$ is as in   \eqref{third}. Taking derivative over
$u_j$ from the both sides of the equality   \eqref{r20},  after that
letting $a_0  \to  \mu$ and  applying $  \phi (y, x) = (-y, x) $ and
$  \phi (  \nu,   \mu) = (-  \nu,   \mu) $,  we have
  \[
  \zeta_j  \left (
  \int_{  \mu}^x  \mathrm{d}  \mathbf{ u}+  \boldsymbol{ u}
  \right)  +  \int_{  \mu}^{x}   \mathrm{d}  r_j   -
  \frac12  \sum_{k=1}^g  \frac{1}{y_k}  \frac{  \partial x_k}{  \partial
u_j}   \frac{y_k-y}{x_k-x}
=
  \zeta_j  \left (
  \boldsymbol{ u}
  \right) -  \frac12  \sum_{k=1}^g  \frac{1}{y_k}  \frac{  \partial x_k}{
 \partial
u_j}   \frac{y_k-  \nu}{x_k-  \mu}.
  \]
Put $x=x_0$. Denoting $P (z) =  \prod_1^g (z-x_j) $ we find
  \begin{eqnarray*}
&&  \sum_{k=1}^g  \frac{1}{y_k}  \frac{  \partial x_k}{  \partial
u_j}   \frac{y_k-y}{x_k-x}=
  \sum_{k=1}^g  \left (  \frac{D_{j} (P (z) ) }{P' (z) }  \Bigg|_{z=x_k}
\right)
  \frac{y_k-y}{x_k-x}  \\=
&&  \sum_{k=0}^g
y_k  \left (  \frac{D_j (R' (z) ) }{R' (z) }  \Bigg|_{z=x_k}  \right)
-  \sum_{k=1}^g
y_k  \left (  \frac{D_{j+1} (P (z) ) }{P' (z) }  \Bigg|_{z=x_k}  \right) .
  \end{eqnarray*}
Hence,  using   \eqref{inversiony} and adding to both sides
$  \sum_{k=1}^g  \int_{a_k}^{x_k}  \mathrm{d}r_j$,  we deduce
  \begin{eqnarray}
&&  \zeta_j  \left (   \int_{  \mu}^{x_0}  \mathrm{d}  \mathbf{
u}+  \boldsymbol{ u}   \right)  +  \int_{  \mu}^{x_0}   \mathrm{d}  r_j
+  \sum_{k=1}^g  \int_{a_k}^{x_k}  \mathrm{d}r_j  -
  \frac12  \sum_{k=0}^g
y_k  \left (  \frac{D_j (R' (z) ) }{R' (z) }  \Bigg|_{z=x_k}  \right)
  \nonumber  \\
&&=
  \zeta_j  \left (
  \boldsymbol{ u}
  \right) +  \sum_{k=1}^g  \int_{a_k}^{x_k}  \mathrm{d}r_j
-  \frac12  \sum_{k=1}^g  \frac{1}{y_k}  \frac{  \partial x_k}{  \partial
u_j}   \frac{y_k-  \nu}{x_k-  \mu}-  \frac12  \wp_{gg, j+1}.  \label{deduce}
  \end{eqnarray}
Now see,  that the left hand side of the   \eqref{deduce}  is
symmetrical in $x_0,  x_1,    \ldots,  x_g$,  while the right hand side
does not depend on $x_0$. So,  it does not depend on any of $x_i$.
We conclude,  that it is a constant  depending only on $  \mu$.
Tending $  \mu  \to a$ and applying the hyperelliptic involution  to
the whole aggregate,  we find this constant to be $0$.   \end{proof}

  \begin{cor} For $ (y, x)   \in V$ and $  \boldsymbol{
  \alpha}=  \int_a^{x}  \mathrm{d}  \mathbf{ u}$ :
  \begin{equation}
  \zeta_{j} (  \boldsymbol{ u}+  \boldsymbol{   \alpha})
-  \zeta_{j} (  \boldsymbol{ u}) -  \zeta_{j} (  \boldsymbol{
  \alpha}) =  \frac{ (-y D_j+  \partial_j)    \mathcal{ P} (x;  \boldsymbol{
u}) } {2  \mathcal{ P} (x;  \boldsymbol{ u}) }
,   \label{stickelberger}   \end{equation}
where $  \partial_j=  \partial/  \partial u_j$.
  \end{cor}

  \begin{proof} To find $  \zeta_{j} (  \boldsymbol{
  \alpha}) $ take the limit
$  \{x_1,   \ldots, x_g  \}  \to  \{a_1,   \ldots, a_g  \}$ in
  \eqref{principalz}. The right hand side of
  \eqref{stickelberger} is obtained by rearranging $
  \frac12  \sum_{k=1}^g  \frac{1}{y_k}  \frac{  \partial x_k}{  \partial
u_j}   \frac{y_k-y}{x_k-x}$.  \end{proof}

  \begin{cor}  The functions
$  \wp_{gggk}$,  for $k=1,   \ldots, g$ are given by   \begin{eqnarray}
  \wp_{gggi}= (6  \wp_{gg}+  \lambda_{2g})   \wp_{gi}+6  \wp_{g, i-1}-2  \wp_{g-1, i}
+  \frac{1}{2}  \delta_{gi}  \lambda_{2g-1}.  \label{wpgggi}
  \end{eqnarray}
  \end{cor}

  \begin{proof} Consider the relation
  \eqref{principalz1}. The differentials $  \mathrm{d}  \zeta_i$,
$i=1,   \ldots, g$ can be presented in the following forms
  \[
-  \mathrm{d}  \zeta_i=  \sum_{k=1}^g  \wp_{ik}  \mathrm{d}u_k=
  \sum_{k=1}^g  \mathrm{d}r_i (x_k) -   \frac12
  \sum_{k=1}^g  \wp_{gg, i+1, k}  \mathrm{d}u_k.
  \]
Put $i=g-1$.
We obtain for each of the $x_k$,  $k=1,   \ldots, g$
  \[
  \left (12x_k^{g+1}+2  \lambda_{2g}x_k^g+
  \lambda_{2g-1}x_k^{g-1}
-4  \sum_{j=1}^g  \wp_{g-1, j}x_k^{j-1}  \right)   \frac{  \mathrm{d}x_k}{ y_k}
=2  \sum_{j=1}^g  \wp_{gggj}x_k^{j-1}  \frac{  \mathrm{d}x_k}{
y_k}.  \]
Applying the formula   \eqref{p} to eliminate the
powers of $x_k$ greater than $g-1$,  and taking into account,  that
the differentials $  \mathrm{d}x_k$ are independent,  we come to
  \[  \sum_{i=1}^g  \left[ (6  \wp_{gg}+  \lambda_{2g})   \wp_{gi}+6  \wp_{g, i-1}
-2  \wp_{g-1, i}
+  \frac{1}{2}  \delta_{gi}  \lambda_{2g-1}  \right]x_k^{i-1}=
  \sum_{j=1}^g  \wp_{gggj}x_k^{j-1}.  \]  \end{proof}

Let us
calculate the difference
$  \frac{  \partial  \wp_{gggk}}{  \partial
u_i}-  \frac{  \partial  \wp_{gggi}}{  \partial u_k}$ according to the
  \eqref{wpgggi}. We obtain
  \begin{cor}
  \begin{equation}
  \wp_{ggk}  \wp_{gi}-  \wp_{ggi}  \wp_{gk}+  \wp_{g, i-1, k}-  \wp_{gi, k-1}=0.
  \label{wp3}
  \end{equation}
  \end{cor}
This means that the $1$--form
$  \sum_{i=1}^g (  \wp_{gg}  \wp_{gi}+  \wp_{g, i-1})   \mathrm{d} u_i$ is
closed. We can rewrite this as $  \mathrm{d}  \boldsymbol{ u}^T  \mathcal{
C}  \boldsymbol{  \wp}$.

Differentiation of   \eqref{wp3} by
$u_g$ yields
  \begin{equation}
  \wp_{gggk}  \wp_{gi}-  \wp_{gggi}  \wp_{gk}+  \wp_{gg, i-1, k}-  \wp_{ggi, k-1}=0.
  \label{i-1, k-1}
  \end{equation}
And the corresponding closed $1$--form is
$  \mathrm{d}   \boldsymbol{ u}^T  \mathcal{ C}  \boldsymbol{   \wp}'$.

  \subsection{Fundamental cubic and quartic
relations}  \label{hyper-Jac} We are going to find the relations
connecting the odd functions $  \wp_{ggi}$ and even functions
$  \wp_{ij}$. These relations take in hyperelliptic theory the place
of the Weierstrass cubic relation   \[ {  \wp'}^2=4  \wp^3-g_2  \wp-g_3,
  \]
for elliptic functions,  which establishes the meromorphic map
between the elliptic Jacobian $  \mathbb{ C}/ (2  \omega, 2  \omega') $ and
the plane cubic.

The theorem below is based
on the property of an Abelian function to be constant if  any
gradient  of it is identically $0$,  or,  if for Abelian
functions $G (  \boldsymbol{ u}) $ and $F (  \boldsymbol{ u}) $ there exist such a
nonzero vector $  \boldsymbol{
  \alpha}  \in   \mathbb{ C}^g$,  that
$  \sum_{i=1}^g  \alpha_i  \frac{  \partial}{  \partial u_i} (G (  \mathbf{
u}) -F (  \boldsymbol{ u}) ) $  vanishes,  then
$G (  \boldsymbol{ u}) -F (  \boldsymbol{ u}) $ is a constant.

  \begin{theorem} The  functions
$  \wp_{ggi}$ and $  \wp_{ik}$ are related by
  \begin{eqnarray}
  \lefteqn{  \wp_{ggi}  \wp_{ggk}=4  \wp_{gg}  \wp_{gi}  \wp_{gk}-
2 (  \wp_{gi}  \wp_{g-1, k}+  \wp_{g, k}  \wp_{g-1, i}) }  \nonumber  \\
&&+
4 (  \wp_{gk}  \wp_{g, i-1}+  \wp_{gi}  \wp_{g, k-1}) +
4  \wp_{k-1, i-1}-2 (  \wp_{k, i-2}+  \wp_{i, k-2})    \nonumber  \\
&&+
  \lambda_{2g}  \wp_{gk}  \wp_{gi}+  \frac{  \lambda_{2g-1}}2
 (  \delta_{ig}  \wp_{kg}+  \delta_{kg}  \wp_{ig}) +c_{ (i, k) },
  \label{product3}
  \end{eqnarray}
where
  \begin{equation}c_{ (i, k) }=  \lambda_{2i-2}  \delta_{ik}
+  \frac12 (  \lambda_{2i-1}  \delta_{k, i+1}
+  \lambda_{2k-1}  \delta_{i, k+1}) .
  \label{cij}
  \end{equation}
  \end{theorem}

  \begin{proof}
We are looking for such a function $G (  \boldsymbol{ u}) $ that
$  \frac{  \partial}{  \partial u_g} (  \wp_{ggi}  \wp_{ggk}-G) $ is $0$.
Direct check using   \eqref{wp3} shows that
  \begin{multline*}
  \frac{  \partial}{  \partial u_g} (  \wp_{ggi}  \wp_{ggk}- (4  \wp_{gg}  \wp_{gi}  \wp_{gk}-
2 (  \wp_{gi}  \wp_{g-1, k}+  \wp_{g, k}  \wp_{g-1, i})   \\
+
4 (  \wp_{gk}  \wp_{g, i-1}+  \wp_{gi}  \wp_{g, k-1}) +
4  \wp_{k-1, i-1}-2 (  \wp_{k, i-2}+  \wp_{i, k-2})    \\
+
  \lambda_{2g}  \wp_{gk}  \wp_{gi}+  \frac{  \lambda_{2g-1}}2
 (  \delta_{ig}  \wp_{kg}+  \delta_{kg}  \wp_{ig}) ) ) =0.
  \end{multline*}
It remains to determine $c_{ij}$.  From   \eqref{principium},  we
conclude,  that $c_{ (i, k) }$  for $k=i$ is equal to
$  \lambda_{2i-2}$,  for $k=i+1$ to $  \frac12{  \lambda_{2i-1}}$,
otherwise $0$. So $c_{ij}$ is given by   \eqref{cij}.  \end{proof}

\noindent Consider $  \mathbb{ C}^{g+ \frac{g (g+1) }{2}}$ with
coordinates $ (\boldsymbol{ z}, p=\{ p_{i, j}  \}_{i, j=1
\ldots g}) $ with $  \boldsymbol{ z}^T= (z_1, \ldots, z_g)$ and
 $p_{ij}=p_{ji}$,  then we
have
  \begin{cor} The map   \[ \varphi:
  \mathrm{Jac} (V)   \backslash (  \sigma)   \to   \mathbb{
C}^{g+  \frac{g (g+1) }{2}},   \quad   \varphi (  \boldsymbol{
u}) = (  \boldsymbol{   \wp}' (  \boldsymbol{ u}) ,    \Pi (  \boldsymbol{ u}) ) ,
  \]
where $  \Pi=  \{  \wp_{ij}  \}_{i, j=1,   \ldots, g}$,  is  meromorphic
embedding.

The image
$  \varphi (  \mathrm{Jac} (V)   \backslash (  \sigma) )   \subset
   \mathbb{ C}^{g+  \frac{g (g+1) }{2}}$ is the intersection of
$  \frac{g (g+1) }{2}$ cubics,  induced by   \eqref{product3}.
  \end{cor}
$ (  \sigma) $ denotes the divisor of $0$'s of $  \sigma$.

Consider projection   \[
  \pi:  \mathbb{ C}^{  \frac{g+g (g+1) }{2}}  \to   \mathbb{
C}^{  \frac{g (g+1) }{2}},   \quad   \pi (  \boldsymbol{ z}, p) =p.
  \]
  \begin{cor}
The restriction  $  \pi  \circ  \varphi$ is the meromorphic embedding of
the Kummer variety
$  \mathrm{Kum} (V) =
(  \mathrm{Jac} (V)   \backslash (  \sigma) ) /  \pm$  into
$  \mathbb{ C}^{  \frac{g (g+1) }{2}}$. The image
$  \pi (  \varphi (  \mathrm{Jac} (V)   \backslash (  \sigma) ) )
\linebreak[3]  \subset   \mathbb{ C}^{  \frac{g (g+1) }{2}}$ is
the intersection of quartics,  induced by \begin{equation} (
 \wp_{ggi}  \wp_{ggj})  (  \wp_{ggk}  \wp_{ggl}) - (  \wp_{ggi}
\wp_{ggk})  (  \wp_{ggj}  \wp_{ggl}) =0, \label{Kijkl}
  \end{equation} where the parentheses mean,  that substitutions
by   \eqref{product3} are made before expanding.  \end{cor} The
quartics   \eqref{Kijkl}  have no analogue in the elliptic theory.
The first example is given by genus $2$, where the celebrated
Kummer surface   \cite{hu94} appears.

  \subsection{Analysis of fundamental relations}  \label{hyper-Kum}
Let us take a second look at the fundamental cubics
  \eqref{product3} and quartics   \eqref{Kijkl}.

  \subsubsection{Sylvester's identity}
For any matrix $K$ of entries $k_{ij}$ with $i, j=1,   \ldots, N$ we
introduce the symbol $K[{}^{i_1}_{j_1}  \cdots{}^{i_m}_{j_n}]$ to
denote the $m  \times n$ submatrix:
  \[
K[{}^{i_1}_{j_1}  \cdots{}^{i_m}_{j_n}]
=  \{k_{i_k, j_l}  \}_{k=1,   \ldots, m;  \, l=1,   \ldots, n}
  \]
for subsets of rows $i_k$ and columns $j_l$.

We will need here the {  \em Sylvester's
identity}  (see,  for instance~  \cite{hj86}) . Let us fix a subset of
indices $  \boldsymbol{  \alpha}=  \{i_1,   \ldots, i_k  \}$,  and make up the
$N-k  \times N-k$ matrix $S (K,   \boldsymbol{  \alpha}) $ assuming that
  \[
S (K,   \boldsymbol{  \alpha}) _{  \mu,   \nu}=  \det
K[{}^{  \mu,   \boldsymbol{  \alpha}}_{  \nu,   \boldsymbol{  \alpha}}]
  \] and
$  \mu,   \nu$ are not in $  \boldsymbol{  \alpha}$,  then
  \begin{equation}
  \det S (K,   \boldsymbol{  \alpha}) =  \det
K[{}^{  \boldsymbol{  \alpha}}_{  \boldsymbol{  \alpha}}]^{ (N-k-1) }
 \det K .
  \label{Sylvester}   \end{equation}

  \subsubsection{Determinantal form}

We introduce  (cf.   \cite{le95})  new functions  $h_{ik}$ defined by
the formula   \begin{eqnarray} h_{ik}&=&4  \wp_{i-1, k-1} -2
  \wp_{k, i-2} -2  \wp_{i, k-2}   \nonumber  \\ &+&
  \frac{1}{2}  \left (  \delta_{ik} (  \lambda_{2i-2}+  \lambda_{2k-2})
+  \delta_{k, i+1}  \lambda_{2i-1}
+  \delta_{i, k+1}  \lambda_{2k-1}  \right) ,
  \label{variables}
  \end{eqnarray}
where the indices
$i, k  \in 1,   \ldots, g+2$.
We assume that $  \wp_{nm}=0$ if $n$ or $m$ is $<1$ and
$  \wp_{nm}=0$ if $n$ or $m$ is $>g$. It is evident that
$h_{ij}=h_{ji}$. We shall denote the matrix of $h_{ik}$ by $H$.

The map   \eqref{variables} from $  \wp$'s and $  \lambda$'s to $h$'s
respects the grading
  \[  \mathrm{deg}  \;h_{ij}=i+j,   \qquad
  \mathrm{deg}  \;  \wp_{ij}=i+j+2,   \qquad
  \mathrm{deg}  \;  \lambda_{i}=i+2,   \] and on a fixed level $L$
  \eqref{variables} is
linear and invertible. From the definition follows
  \[
  \sum_{i=1}^{L-1}h_{i, L-i}=  \lambda_{L-2}   \Rightarrow   \boldsymbol{
X}^T H  \boldsymbol{ X}=  \sum_{i=0}^{2g+2}  \lambda_{i}x^{i}
  \]
for $  \boldsymbol{ X}^T= (1, x,   \ldots, x^{g+1}) $ with
arbitrary $x  \in  \mathbb{ C}$. Moreover, for any roots $x_r$ and
$x_s$ of the equation $  \sum_{j=1}^{g+2} h_{g+2, j}x^{j-1}=0$ we
have  (cf.   \eqref{principal11})  $ y_ry_s =  \boldsymbol{ X}_r^T
H  \boldsymbol{ X}_s$.

From   \eqref{variables} we have
\begin{gather}
-2  \wp_{ggi}=  \tfrac{  \partial}{  \partial
u_g}h_{g+2, i}=  \tfrac{  \partial}{  \partial
u_i}h_{g+2, g}=-  \tfrac12  \tfrac{  \partial}{  \partial
u_i}h_{g+1, g+1},   \notag  \\
2 ( \wp_{gi, k-1}- \wp_{g, i-1, k}) =  \tfrac{  \partial}{
  \partial u_k}h_{g+2, i-1}- \tfrac{  \partial}{  \partial
u_i}h_{g+2, k-1}= \tfrac12 \tfrac{  \partial}{  \partial
u_k}h_{g+1, k}
-\tfrac12  \tfrac{  \partial}{  \partial
u_i}h_{g+2, i}, \notag  \\
  \intertext{$  \ldots$ etc.,    \,   and  (see   \eqref{wpgggi}) :}
-2  \wp_{gggi}= \tfrac{ \partial^2}{\partial
u_g^2}h_{g+2, i}=
-  \det H[{}^{i, }_{g+1 }{}^{g+1 }_{g+2}] -
  \det H[{}^{i-1, }_{g+1,}{}^{g+2 }_{g+2}] -
  \det H[{}^{i, }_{g,}{}^{g+2}_{g+2}].\label{dg_H}
\end{gather}
Using \eqref{variables},  we write  \eqref{product3} in
more effective form:
\begin{eqnarray}
4  \wp_{ggi}  \wp_{ggk}=  \tfrac{  \partial}{  \partial
u_g}h_{g+2, i}  \tfrac{  \partial}{  \partial
u_g}h_{g+2, k}=- \det
H[{}^{i, }_{k, }{}^{g+1, }_{g+1, }{}^{g+2}_{g+2}]
  \label{wpggiwpggk}
  \end{eqnarray}

Consider,  as an example,  the case of genus $1$. We define on the
Jacobian of a curve   \[ y^2=  \lambda_4 x^4+  \lambda_3 x^3 +  \lambda_2
x^2+  \lambda_1 x+  \lambda_0    \] the Kleinian functions:
$  \sigma_{K} (u_1) $ with expansion $u_1+  \ldots$,  its second  and
third logarithmic derivatives $-  \wp_{11}$ and $-  \wp_{111}$. By
  \eqref{wpggiwpggk} and following the definition   \eqref{variables}
  \[
  -4\wp_{111}^2=\det H
  \left[{}_{1, 2, 3}^{1, 2, 3}  \right]= \det  \left (
  \begin{array}{ccc}   \lambda_0&   \tfrac{1}{2}  \lambda_1&-2
  \wp_{11}  \\ \tfrac12 \lambda_1&4  \wp_{11}+  \lambda_2&
  \tfrac12  \lambda_3  \\ -2  \wp_{11}&  \tfrac12
\lambda_3&  \lambda_4 \end{array} \right) ; \] the determinant
 expands as:  \[ \wp_{111}^2=4  \wp_{11}^3+  \lambda_2
  \wp_{11}^2+ \wp_{11}  \frac{  \lambda_1   \lambda_3-4  \lambda_4
  \lambda_0}{4}+ \frac{  \lambda_0   \lambda_3^2+  \lambda_4 (
 \lambda_1^2-4  \lambda_2  \lambda_0) }{16}, \] and the
  \eqref{dg_H},  in complete accordance, gives \[ \wp_{1111}=6
  \wp_{11}^2 +  \lambda_2   \wp_{11}+ \frac{  \lambda_1
  \lambda_3-4  \lambda_4  \lambda_0}{8} \] These equations show
that $  \sigma_K$ differs by $  \mathrm{exp} (-  \frac1{12}
\lambda_2 u_1^2) $ from standard Weierstrass $  \sigma_W$ built by
the invariants $g_2=  \lambda_4  \lambda_0+  \frac1{12}
\lambda_2^2-  \frac14  \lambda_3  \lambda_1$ and $g_3=  \det
\left (  \begin{smallmatrix} \lambda_0&  \frac14  \lambda_1&
  \frac16  \lambda_2  \\ \frac14  \lambda_1&  \frac16  \lambda_2&
  \frac14  \lambda_3  \\ \frac16  \lambda_2&  \frac14  \lambda_3&
  \lambda_4 \end{smallmatrix}  \right) $  (see,  e.g.
  \cite{ba55, ww73}) .

Further,
we find,  that $  \mathrm{rank}  \; H=3$ in generic point of Jacobian,
$  \mathrm{rank}  \;  H=2$ in halfperiods.  At $u_1=0$,  where
$  \sigma_K$ has is $0$ of order $1$,  we have $  \mathrm{rank}  \;
  \sigma_K^2 H=3$.

Concerning the general case,  on the ground of   \eqref{wpggiwpggk},
we prove the following:
  \begin{theorem}
$  \mathrm{rank}  \;H=3$ in generic point $  \in  \mathrm{Jac} (V) $  and
$  \mathrm{rank}  \;H=2$ in the halfperiods.
$  \mathrm{rank}  \;  \sigma (  \boldsymbol{ u}) ^2 H=3$ in generic
point $  \in  (  \sigma) $ and
$  \mathrm{rank}  \;  \sigma (  \boldsymbol{ u}) ^2 H$ in the  points
of $ (  \sigma) _{  \mathrm{sing}}$.
  \end{theorem}
Here $ (  \sigma)   \subset   \mathrm{Jac} (V) $ denotes the divisor of
$0$'s of $  \sigma (  \boldsymbol{ u}) $. The
$ (  \sigma) _{  \mathrm{sing}}  \subset  (  \sigma) $ is the so-called
singular set of $ (  \sigma) $. $ (  \sigma) _{  \mathrm{sing}}$ is the set
of points where $  \sigma$ vanishes and all its first partial
derivatives vanish.  $ (  \sigma) _{  \mathrm{sing}}$ is known  (see
  \cite{fa73} and references therein)  to be a subset of dimension
$g-3$ in hyperelliptic Jacobians of $g>3$,  for  genus $2$ it is
empty and consists of single point for $g=3$. Generally,  the
points of $ (  \sigma) _{  \mathrm{sing}}$ are presented by
$  \{ (y_1, x_1) ,   \ldots,  (y_{g-3}  \, , x_{g-3}  \, )   \}  \in (V) ^{g-3}$ such that
for all $i  \neq j  \in 1,   \ldots,  g-3, $ $  \phi (y_i, x_i)   \neq (y_j, x_j) $.

  \begin{proof} Consider the Sylvester's
matrix $S=S   \left (H
 [{}^{i, j, g+1, g+2}_{k, l, g+1, g+2}],   \{g+1, g+2  \}
  \right) $.
By   \eqref{wpggiwpggk} we have
$S=-4  \left (   \begin{array}{ll}
  \wp_{ggi}  \wp_{ggk}&  \wp_{ggi}  \wp_{ggl}  \\
  \wp_{ggj}  \wp_{ggk}&  \wp_{ggj}  \wp_{ggl}
  \end{array}  \right) $
and $  \det S=0$,  so by   \eqref{Sylvester} we see,  that
$  \det H  [{}^{i, j, g+1, g+2}_{k, l, g+1, g+2} ]
  \det H  [{}^{g+1, g+2}_{g+1, g+2} ]$ vanishes
identically. As  $  \det H  [{}^{g+1, g+2}_{g+1, g+2} ]=
  \lambda_{2g+2} (4  \wp_{gg}+  \lambda_{2g}) -  \frac14  \lambda_{2g+1}^2$
is not an identical $0$,  we infer that
  \begin{equation}
  \det H  \left[{}^{i, j, g+1, g+2}_{k, l, g+1, g+2}  \right]=0.
  \label{subHdet}
  \end{equation}
Remark,  that this equation is actually the   \eqref{Kijkl} rewritten
in terms of $h$'s. Now from the   \eqref{subHdet},  putting $j=l=g$,
we obtain for any $i, k$,  except for such $  \boldsymbol{ u}$,  that
$H  \left[{}^{g, g+1, g+2}_{g, g+1, g+2}  \right]$ becomes degenerate,  and
those  where the entries become singular i.e.
$  \boldsymbol{u}  \in (  \sigma) $,    \begin{equation}
h_{ik}= (h_{i, g}, h_{i, g+1}, h_{i, g+2})
  \left (H  \left[{}^{g, g+1, g+2}_{g, g+1, g+2}  \right]  \right) ^{-1}
  \left (  \begin{array}{l}
h_{k, g}  \\
h_{k, g+1}  \\
h_{k, g+2}
  \end{array}
  \right) .  \label{hik}
  \end{equation}
This leads to the skeleton decomposition of the matrix $H$
  \begin{equation}
H=H  \left[{}^{1,   \, {  \displaystyle   \ldots}  \, , g+2}_{g, g+1, g+2}
 \right]
  \left (H  \left[{}^{g, g+1, g+2}_{g, g+1, g+2}  \right]  \right) ^{-1}
H  \left[{}^{g, g+1, g+2}_{1,   \, {  \displaystyle   \ldots}  \, , g+2}
 \right],
  \label{skeleton}
  \end{equation}
which shows,  that in generic point of $  \mathrm{Jac} (V) $ rank
of $H$ equals $3$.

Consider the case $  \det
H  [{}^{g, g+1, g+2}_{g, g+1, g+2}]=0$. As by
  \eqref{wpggiwpggk} we have $   \det
H [{}^{g, g+1, g+2}_{g, g+1, g+2}]= 4 \wp_{ggg}^2$,  this may
happen only iff $  \boldsymbol{u}$ is a halfperiod.  And therefore
  we have instead of   \eqref{subHdet} the equalities $H
[{}^{i, g+1, g+2}_{k, g+1, g+2}]=0$ and consequently
in halfperiods matrix $H$ is decomposed as
  \[
H=H  \left[{}^{1,   \, {  \displaystyle   \ldots}  \, , g+2}_{g+1, g+2}
 \right]
  \left (H  \left[{}^{g+1, g+2}_{g+1, g+2}  \right]  \right) ^{-1}
H  \left[{}^{g+1, g+2}_{1,   \, {  \displaystyle   \ldots}  \, , g+2}
 \right],\]
having the rank $2$.

Next,  consider $  \sigma (  \boldsymbol{ u}) ^2 H$ at the
$  \boldsymbol{ u}  \in (  \sigma) $. We have
$  \sigma (  \boldsymbol{ u}) ^2
h_{i, k}=4  \sigma_{i-1}  \sigma_{k-1}-2  \sigma_{i}  \sigma_{k-2}
-2  \sigma_{i-2}  \sigma_{k}, $  where
$  \sigma_{i}=  \frac{  \partial}{  \partial u_i}  \sigma (  \boldsymbol{
u}) $,  and,  consequently,  the
decomposition
  \[
   \sigma (  \boldsymbol{ u}) ^2 H|_{  \boldsymbol{
u}  \in (  \sigma) }=2 (  \boldsymbol{ s}_1,    \boldsymbol{ s}_2,
  \boldsymbol{ s}_3)    \left (   \begin{array}{rrr} 0&0&-1  \\ 0&2&0  \\
-1&0&0   \end{array}   \right)    \left (   \begin{array}{r}   \boldsymbol{
s}_1^T  \\   \boldsymbol{ s}_2^T  \\   \boldsymbol{ s}_3^T   \end{array}
  \right) ,    \] where $  \boldsymbol{
s}_1= (  \sigma_1,   \ldots,   \sigma_g, 0, 0) ^T$,  $  \boldsymbol{
s}_2= (0,   \sigma_1,   \ldots,   \sigma_g, 0) ^T$  and $  \boldsymbol{
s}_3= (0, 0,   \sigma_1,   \ldots,   \sigma_g) ^T$. We infer,  that
$  \mathrm{rank} (  \sigma (  \boldsymbol{ u}) ^2 H) $ is $3$ in generic
point of $ (  \sigma) $,  and
becomes $0$ only when $  \sigma_1=  \ldots=  \sigma_g=0$,   is in the
points $  \in  (  \sigma) _{  \mathrm{sing}}$,  while no other values are
possible.
  \end{proof}

{  \em Conclusion}. The map   \begin{align*}
h:  \boldsymbol{ u}  \mapsto&
  \{4  \sigma_{i-1}  \,   \sigma_{k-1}-2  \sigma_{i}  \,   \sigma_{k-2}
-2  \sigma_{i-2}  \,   \sigma_{k}  \\&-  \sigma
 (4  \sigma_{i-1, k-1}  \, -2  \sigma_{i, k-2}  \,
-2  \sigma_{i-2, k}) +
  \tfrac{1}{2}  \sigma^2 (  \delta_{ik} (  \lambda_{2i-2}+  \lambda_{2k-2})   \\
&+  \delta_{k, i+1}  \lambda_{2i-1}
+  \delta_{i, k+1}  \lambda_{2k-1})   \}_{i, k  \in 1,   \ldots,  g+2},
   \end{align*}  induced by
  \eqref{variables} establish the meromorphic map of the
$  \big (  \mathrm{Jac} (V)   \backslash (  \sigma) _{  \mathrm{sing}}  \big) /  \pm
$ into the space $Q_3$
of complex symmetric $ (g+2)   \times (g+2) $ matrices of
$  \mathrm{rank}$ not greater than $3$.

We give the example of genus   $2$ with $  \lambda_6=0$ and
$  \lambda_5=4$:

  \begin{equation}
H=  \left (  \begin{array}{cccc}  \lambda_0&  \frac{1}{2}  \lambda_1&-2
 \wp_{11}&-2  \wp_{12}  \\
  \frac{1}{2}  \lambda_1&  \lambda_2+4  \wp_{11}&  \frac{1}{2}  \lambda_3-
2  \wp_{12}&-2  \wp_{22}  \\-2  \wp_{11}&  \frac{1}{2}  \lambda_3-2
 \wp_{12}&  \lambda_4+4  \wp_{22}&2  \\
-2  \wp_{12}&-2  \wp_{22}&2&0  \end{array}  \right) .
  \label{kum}  \end{equation}
In this case $ (  \sigma) _{  \mathrm{sing}}=  \{  \varnothing  \}$,
so the Kummer surface in $  \mathbb{ C}  \mathbb{ P}^3$ with coordinates
 \newline
$ (X_0,  X_1,  X_2,  X _3)  =  (  \sigma^2,   \sigma^2  \wp_{11},   \sigma^2  \wp_{12},   \sigma^2  \wp_{22}) $ is
defined by the equation $  \det  \sigma^2 H=0$.

  \subsubsection{Extended cubic relation}
The extension   \cite{le95} of   \eqref{wpggiwpggk} is given by
  \begin{theorem}
  \begin{eqnarray}
  \mathbf{ R}^T  \boldsymbol{   \pi}_{jl}  \boldsymbol{   \pi}_{ik}^T
 \mathbf{
S}=  \frac{1}{4}  \det   \left (  \begin{array}{cc} H
  \left[{}^i_j{}^k_l{}^{g+1}_{g+1}{}^{g+2}_{g+2}  \right]
&  \mathbf{ S}  \\  \mathbf{ R}^T&0
  \end{array}  \right) ,   \label{bakergen}
  \end{eqnarray}
where $  \mathbf{ R},   \,    \mathbf{ S}  \in   \mathbb{  C}^4$ are
arbitrary vectors and
  \[  \boldsymbol{   \pi}_{ik}=  \left (
  \begin{array}{c} -  \wp_{ggk}  \\
  \wp_{ggi}  \\
  \wp_{g, i, k-1}-  \wp_{g, i-1, k}  \\
  \wp_{g-1, i, k-1}-
  \wp_{g-1, k, i-1}+
  \wp_{g, k, i-2}-
  \wp_{g, i, k-2}
  \end{array}
  \right)   \]
  \end{theorem}

  \begin{proof}
Vectors $  \boldsymbol{ {  \tilde  \pi}}=  \boldsymbol{   \pi}_{ik}$ and
$  \boldsymbol{   \pi}=  \boldsymbol{   \pi}_{jl}$ solve the equations    \[
H  \left[{}^i_j{}^k_l{}^{g+1}_{g+1}{}^{g+2}_{g+2}  \right]  \boldsymbol{
  \pi}=0;   \quad   \boldsymbol{
{  \tilde  \pi}}^T{H}  \left[{}^i_j{}^k_l{}^{g+1}_{g+1}{}^{g+2}_{g+2}
  \right]=0.
  \]
The theorem follows.
  \end{proof}

The case of genus $2$,  when $  \boldsymbol{
  \pi}_{21}= (-  \wp_{222},   \wp_{221}, -  \wp_{211},   \wp_{111}) ^T$
exhausts all the possible $  \wp_{ijk}$--functions,  the relation
  \eqref{bakergen} was thoroughly studied by Baker   \cite{ba07}.

  \section{Applications}

  \subsection{Matrix realization of hyperelliptic Kummer varieties}
Here we present the explicit matrix realization  (see   \cite{bel96})
of hyperelliptic Jacobians $  \mathrm{Jac} (V) $  and Kummer varieties
$  \mathrm{Kum} (V) $ of the curves $V$ with the fixed branching point
$e_{2g+2}=a=  \infty$.  Our approach is based on the results of
Section   \ref{hyper-Kum}.

Let us consider the space $  \mathcal{ H}$ of complex symmetric
$ (g+2)   \times (g+2) $--matrices  $  \mathrm{ H}=  \{  \mathrm{
h}_{k, s}  \}$,   with $  \mathrm{ h}_{g+2, g+2}=0$ and $  \mathrm{
h}_{g+1, g+2}=2$.  Let us put in correspondence  to $  \mathrm{
H}  \in   \mathcal{ H}$  a symmetric $g  \times g$--matrix $  \mathrm{
A} (  \mathrm{ H}) $,  with entries $a_{k, s}=  \det   \mathrm{ H}
  \left[{}_{s, g+1, g+2}^{k, g+1, g+2}  \right]$.

From the Sylvester's identity   \eqref{Sylvester} follows that
 rank of the matrix $  \mathrm{ H}  \in   \mathcal{ H}$
does not exceed $3$ if and only if rank of the  matrix
$  \mathrm{ A} (  \mathrm{ H}) $ does not exceed $1$.

Let us put  $K  \mathcal{ H}=   \left  \{  \mathrm{ H}  \in   \mathcal{ H}:
  \mathrm{rank}  \mathrm{ H}  \leq 3  \right  \} $. For each complex
symmetric $g  \times g$--matrix $  \mathrm{ A}=  \{a_{k, s}  \}$ of rank not
greater $1$,  there exists,  defined up to sign,
a $g$--dimensional column vector $
  \mathbf{  z}=  \mathbf{  z} (  \mathrm{ A}) $,  such that
$  \mathrm{ A}=-4  \mathbf{  z}  \cdot   \mathbf{  z}^T$.

Let us introduce vectors
$  \mathbf{  h}_k=  \{  \mathrm{ h}_{k, s};  \;s=1,   \ldots, g   \}  \,
 \in   \mathbb{
C}^g$.

  \begin{lemma}  \label{geom-2}
  Map
  \begin{eqnarray*}
&  \gamma: K  \mathcal{ H}   \to  (  \mathbb{ C}^g/  \pm)   \times
 \mathbb{ C}^g  \times  \mathbb{
C}^g  \times  \mathbb{ C}^1  \\
&  \gamma (  \mathrm{ H}) =
-  \left (  \mathbf{  z}  \left (  \mathrm{ A} (  \mathrm{ H})   \right) ,
   \mathbf{
h}_{g+1},    \mathbf{  h}_{g+2},   \mathrm{h}_{g+1, g+1}  \right)
  \end{eqnarray*}
 is a homeomorphism.
  \end{lemma}

  \begin{proof} follows from the relation:
  \[
4   \hat{  \mathrm{ H}}=4   \mathbf{  z}  \cdot   \mathbf{  z}^T+2
  \left (  \mathbf{  h}_{g+2}  \mathbf{  h}_{g+1}^T+  \mathbf{  h}_{g+1}
  \mathbf{  h}_{g+2}^T  \right) -  \mathrm{h}_{g+1, g+1}  \mathbf{  h}_{g+2}  \mathbf{  h}_{g+2}^T
  \]
where $  \hat{  \mathrm{ H}}$ is the matrix composed of the column
vectors $  \mathbf{  h}_k,    \,  k=1,   \ldots, g$,  and $  \mathbf{  z}= (  \mathbf{
z}  \left (  \mathrm{ A} (  \mathrm{ H})   \right) $.
  \end{proof}

Let us introduce  the $2$--sheeted ramified
 covering
$  \pi:J  \mathcal{ H}  \to K  \mathcal{ H}$,   which the
covering $  \mathbb{ C}^g  \to (  \mathbb{ C}^g/  \pm) $ induces by the map
 $  \gamma$.

  \begin{cor}
 $  \hat   \gamma: J  \mathcal{ H}   \cong   \mathbb{ C}^{3g+1}$.
  \end{cor}

Now let us consider the  universal space
$W_g$ of $g$--th
symmetric powers of hyperelliptic curves
  \[V=
  \left  \{  (y, x)    \in   \mathbb{ C}^2:   \, y^2=4 x^{2g+1} +
  \sum_{k=0}^{2g}  \lambda_{2g-k}x^{2g-k}  \right  \}   \]  as an algebraic
subvariety in $ (  \mathbb{ C}^2) ^g  \times
  \mathbb{ C}^{2g+1}$ with coordinates
  \[  \left  \{
  \left ( (y_1, x_1) ,   \ldots,  (y_g, x_g)   \right) ,   \;  \lambda_{2g},   \ldots,
  \lambda_0   \right  \},   \] where $ (  \mathbb{ C}^2) ^g$ is  $g$--th
symmetric power of the space $  \mathbb{ C}^2$.

Let us define the map   \[  \lambda: J  \mathcal{ H}  \cong
 \mathbb{ C}^{3g+1}  \to
 (  \mathbb{ C}^2) ^g  \times   \mathbb{ C}^{2g+1}  \] in the following way:
  \begin{itemize}
  \item for
$  \boldsymbol{ G}= (  \mathbf{  z},   \mathbf{  h}_{g+1},   \mathbf{
h}_{g+2},   \mathrm{h}_{g+1, g+1})    \in   \mathbb{ C}^{3g+1}$ construct
by Lemma
  \ref{geom-2} the matrix $  \pi (  \boldsymbol{ G}) =  \mathrm{
H}=  \{  \mathrm{ h}_{k, s}  \}  \in K  \mathcal{ H}$
  \item  put
  \[  \lambda (  \boldsymbol{ G}) =  \{ (y_k, x_k) ,
  \lambda_r;  \;k=1,   \ldots, g,   \,  r=0,   \ldots,  2g,
  \}  \]
where $  \{x_1,    \ldots,  x_g  \}$ is the set of roots of the equation
$2 x^g+   \mathbf{  h}_{g+2}^T   \mathbf{ X}=0$,  and
$y_k=  \mathbf{  z}^T   \mathbf{ X}_k$,  and
$  \lambda_r=  \sum_{i+j=r+2}   \mathrm{h}_{i, j}$.
  \end{itemize}
Here
$  \mathbf{ X}_k= (1, x_k,   \ldots, x_k^{g-1}) ^T$.

  \begin{theorem}. Map  $  \lambda$ induces map
$J  \mathcal{ H}  \cong   \mathbb{ C}^{3g+1}  \to W_g$ .  \end{theorem}

  \begin{proof} Direct check shows,  that  the identity is valid
  \[
  \mathbf{ X}_k^T   \mathrm{ A}
  \mathbf{ X}_s+4  \sum_{i, j=1}^{g+2}
  \mathrm{h}_{i, j}x_{k}^{i-1}x_s^{j-1}=0,    \]
where $  \mathrm{
A}=  \mathrm{ A} (  \mathrm{ H}) $  and $  \mathrm{ H}=  \pi (  \boldsymbol{
 G}) $.  Putting $k=s$ and using $  \mathrm{ A}=4   \mathbf{  z}  \cdot
  \mathbf{  z}^T$,  we have $y_k^2=4
x_k^{2g+1}+  \sum_{s=0}^{2g}  \lambda_{2g-s}x_k^{2g-s}$.  \end{proof}

Now it is all ready to give the description of our realization
of varieties $T^g=  \mathrm{Jac} (V) $ and
$K^g=  \mathrm{Kum} (V) $ of the hyperelliptic curves.

For each nonsingular curve
$V=   \left  \{
 (y, x) ,  y^2=4
x^{2g+1} +   \sum_{s=0}^{2g}  \lambda_{2g-s}x^{2g-s}  \right  \} $
define the map
  \[  \gamma:  \;T^g  \backslash (  \sigma)   \to   \mathcal{
H}:  \gamma (u) =  \mathrm{ H}=  \{  \mathrm{h}_{k, s}  \},   \] where
$  \mathrm{h}_{k, s}=4  \wp_{k-1, s-1}-2 (  \wp_{s, k-2}+  \wp_{s-2, k})
+  \frac12
[  \delta_{ks} (  \lambda_{2s-2}+  \lambda_{2k-2}) +  \delta_{k+1, s}
  \lambda_{2k-1}+  \delta_{k, s+1}  \lambda_{2s-1}]$.

  \begin{theorem} The map  $  \gamma$  induces map
$T^g  \backslash (  \sigma)   \to K  \mathcal{ H}$,  such that
$  \wp_{ggk}  \wp_{ggs}=  \dfrac14 a_{ks} (  \gamma (u) ) $,   i.e
$  \gamma$  is lifted to
  \[
{  \tilde   \gamma}:T^g  \backslash (  \sigma)   \to J  \mathcal{ H}  \cong   \mathbb{
C}^{3g+1}   \;   \text{with }  \;   \mathbf{  z}= (  \wp_{gg1},   \ldots,
 \wp_{ggg} ) ^T.    \]
Composition of maps $  \lambda{  \tilde
  \gamma}:  \;T^g  \backslash (  \sigma)   \to W_g$  defines
the inversion of the Abel map $  \mathfrak{ A}:  \;  (V) ^g  \to T^g$ and,
therefore,  the map $  \tilde  \gamma$ is an embedding.
  \label{geom-3}
  \end{theorem}

So we have obtained the explicit realization of the Kummer variety
$T^g  \backslash (  \sigma) /  \pm$  of the hyperelliptic
curve  $V$ of genus $g$ as a subvariety in the
variety of matrices $K  \mathcal{ H}$.  As a consequence of the  Theorem
  \ref{geom-3},  particularly,  follows the new proof of the theorem
by B.A. Dubrovin and S.P. Novikov  about rationality of the
universal space of the Jacobians of hyperelliptic curves
$V$ of genus  $g$ with the fixed branching point $e_{2g+2}=  \infty$
  \cite{dn74}.

  \subsection{Hyperelliptic $  \Phi$--function}
 In this section we construct the linear
differential operators,  for which the hyperelliptic curve
$V (y, x) $ is the spectral variety.

  \begin{definition}
$  \Phi$--function of the curve $V (y, x) $ with fixed
point $a$
  \begin{eqnarray*}
&  \Phi:  \mathbb{ C}  \times  \mathrm{Jac} (V)   \times V
 \to   \mathbb{ C}  \\
&  \Phi (u_0,   \boldsymbol{
u}; (y, x) ) =  \dfrac{  \sigma (  \boldsymbol{
  \alpha}-  \boldsymbol{ u}) }{  \sigma (  \boldsymbol{
  \alpha})   \sigma (  \boldsymbol{ u}) }  \exp (-  \frac{1}{2}y u_0+
 \boldsymbol{
  \zeta}^T (  \boldsymbol{
  \alpha})   \boldsymbol{ u}) ,
  \end{eqnarray*}
where $  \boldsymbol{
  \zeta}^T (  \boldsymbol{
  \alpha}) = (
  \zeta_1 (  \boldsymbol{
  \alpha}) ,   \ldots,
  \zeta_g (  \boldsymbol{
  \alpha}) ) $ and $ (y, x)   \in V$,
$  \boldsymbol{ u}$ and $  \boldsymbol{
  \alpha}=  \int_a^x  \mathrm{d}  \mathbf{ u}$ $  \in   \mathrm{Jac} (V) $.
  \end{definition}

Particularly,  $  \Phi (0,   \boldsymbol{
u}; (y, x) ) $ is the Baker function   (see   \cite[page 421]{ba97} and
  \cite{kr77}) .

  \begin{theorem} The function $  \Phi=  \Phi (u_0,   \boldsymbol{
u}; (y, x) ) $ solves the Hill's equation
  \begin{equation}
 (  \partial_g^2-2  \wp_{gg})   \Phi= (x+  \frac{  \lambda_{2g}}{4})   \Phi,
  \label{Hill}  \end{equation}
with respect to $u_g$,  for all $ (y, x)   \in V$.
  \end{theorem}

  \begin{proof} From   \eqref{stickelberger}
  \[
  \partial_g  \Phi=  \frac{y+  \partial_g{  \mathcal{ P}} (x;
\boldsymbol{ u}) } {2
  \mathcal{ P} (x;  \boldsymbol{ u}) }   \Phi,
  \]
where
$  \mathcal{ P} (x;  \boldsymbol{  u}) $ is given by   \eqref{p},
hence:
  \[
  \frac{  \partial_g^2  \Phi}{  \Phi}=  \frac{y^2- (  \partial_g  \mathcal{
P} (x;  \boldsymbol{  u}) ) ^2 +2  \mathcal{ P} (x;  \boldsymbol{
u})   \partial_g^2  \mathcal{ P} (x;  \boldsymbol{ u}) }{4  \mathcal{
P}^2 (x;  \boldsymbol{  u}) }   \]   and by   \eqref{product3} and
  \eqref{wpgggi} we obtain the theorem.  \end{proof}

Let us introduce the vector $  \boldsymbol{   \Psi}= (  \Phi,   \Phi_g) $,
where $  \Phi_g$ stands for $  \partial_g    \Phi$. Then equation
  \eqref{Hill} may be written as
  \begin{equation}
  \partial_g  \boldsymbol{   \Psi}= L_g   \boldsymbol{   \Psi},
  \,   \text{where}   \quad L_g=   \begin{pmatrix}
0&1  \\
x+2  \wp_{gg}+  \frac{  \lambda_{2g}}{4}&0
  \end{pmatrix}.   \label{matrix-Hill}   \end{equation}

In regard of   \eqref{matrix-Hill}  and   \eqref{stickelberger},  it is
natural to introduce the family of $g+1$ operators,  presented by
$2   \times 2$ matrices,    \[  \{ L_0, L_1,   \ldots, L_g  \},    \quad L_k=
  \left (   \begin{array}{rr} V_k&U_k  \\W_k&-V_k   \end{array}   \right)
   \]
and defined by the equalities
  \[
L_k  \boldsymbol{   \Psi}={  \partial_k}   \boldsymbol{
  \Psi},   \quad k  \in 0,   \ldots, g.
  \]
The theory developed in previous sections leads to the following
description of this family of operators.
  \begin{prop} Entries of
the matrices $L_k$ are polynomials in $x$ and $2g$--periodic in
$  \boldsymbol{ u}$:    \begin{gather} L_k=D_k L_0-  \frac12
  \begin{pmatrix} 0&0  \\h_{g+2, k}&0   \end{pmatrix},   \notag  \\
  \intertext{with}
U_0=  \frac12  \sum_{i=1}^{g+2}x^{i-1}h_{g+2, i},   \quad
V_0=-  \frac14  \sum_{i=1}^{g+2}x^{i-1}  \partial_g h_{g+2, i},    \\
  \quad  \text{and}  \quad W_0=  \frac14  \sum_{i=1}^{g+2}x^{i-1}  \det
  \begin{pmatrix} h_{g+1, i}&h_{g+2, g}  \\h_{g+2, i}&h_{g+2,
  g+1}  \notag \end{pmatrix}.  \label{entries} \end{gather} And
the compatibility conditions \[ [L_k, L_i]=  \partial_k L_i -
        \partial_i L_k   \] are satisfied.  \end{prop} Here $D_k$ is
umbral derivative  (see page~  \pageref{umbral_D}) .  Proof is
straightforward due to   \eqref{stickelberger}, \eqref{variables},
  \eqref{dg_H} and   \eqref{wpggiwpggk}.

  \begin{theorem}  \label{the-F} The function
$  \Phi=  \Phi (u_0,   \boldsymbol{ u}; (y, x) ) $ solves the system of
equations   \begin{gather}   \left (  \partial_k   \partial_l
-  \gamma_{kl} (x,   \boldsymbol{ u})   \partial_g +
  \beta_{kl} (x,   \boldsymbol{ u})
  \right)   \Phi=  \tfrac{1}{4}D_{k+l}  \big (f (x)   \big)   \Phi
  \notag  \\ \intertext{with polynomials in $x$} \begin{aligned}
  \gamma_{kl} (x,   \boldsymbol{ u}) &=
  \tfrac{1}{4}   \left[
  \partial_k D_l +
  \partial_l D_k
  \right]  \sum_{i=1}^{g+2}x^{i-1}h_{g+2, i}  \quad  \text{and}  \\
  \beta_{kl} (x,   \boldsymbol{ u}) &=  \tfrac{1}{8}   \left[
 (  \partial_g  \partial_k +h_{g+2, k} )  D_l +
 (  \partial_g  \partial_l +h_{g+2, l} )  D_k
  \right]  \sum_{i=1}^{g+2}x^{i-1}h_{g+2, i}  \\
&-  \frac14   \sum  \limits_{j=k+l+2}^{2g+2}x^{j- (k+l+2) }
  \left[ \left (  \sum  \limits_{  \nu=1}^{k+1} h_{  \nu, j-  \nu}
  \right) + \left (  \sum  \limits_{  \mu=1}^{l+1} h_{j-  \mu,
  \mu}  \right)    \right] \end{aligned}  \notag \end{gather} for
all $k, l   \in 0,   \ldots,  g$ and arbitrary $ (y, x)    \in V$.
  \end{theorem}
Here $f (x) $ is as given in   \eqref{curve} with
$  \lambda_{2g+2}=0$ and $  \lambda_{2g+1}=4$.
  \begin{proof}
Construction of operators $L_k$ yields
   \[
  \Phi_{lk}=  \tfrac12   \left (
  \partial_l U_k+
  \partial_k U_l  \right) {  \Phi_g}+  \left (V_lV_k +  \tfrac12
 (   \partial_l V_k+  \partial_k
V_l+U_k W_l+W_k U_l)   \right)   \Phi.
  \]
To prove the theorem we use   \eqref{entries},  and it only remains
to notice,  that   (cf. Lemma   \ref{geom-2}) :
  \begin{eqnarray*}
&&D_k (V_0) D_l (V_0) +  \tfrac12 D_k (U_0) D_l (W_0)
+  \tfrac12 D_l (U_0) D_k (W_0) =  \\&&
-  \tfrac{1}{16}  \left (  \det
H   \left[{}_{g+1}^{g+1}{}_{g+2}^{g+2}  \right]
  \right)
 (1, x,   \ldots, x^{g+1-k})
H   \left[{}_{k}^{l}{}_{  \ldots}^{  \ldots}{}_{g+2}^{g+2}  \right]
 (1, x,   \ldots, x^{g+1-l}) ^T,
  \end{eqnarray*}
having in mind that $h_{g+2, g+2}=0$ and $h_{g+2, g+1}=2$,  we obtain
the theorem due to properties of matrix $H$.
  \end{proof}

Consider as an example the case of genus $2$.
  \begin{eqnarray*}
&& (  \partial_2^2-2  \wp_{22})
  \Phi=  \tfrac{1}{4} (4 x+  \lambda_4)   \Phi,   \\
&& (  \partial_2  \partial_1+  \tfrac{1}{2}  \wp_{222}  \partial_2
-  \wp_{22} (x+  \wp_{22}+
  \tfrac{1}{4}  \lambda_4)  +2  \wp_{12})
  \Phi=  \tfrac{1}{4} (4 x^2+  \lambda_4 x +  \lambda_3)   \Phi,    \\
&& (  \partial_1^2+  \wp_{122}  \partial_2
-2  \wp_{12} (x+
  \wp_{22}+  \tfrac{1}{4}  \lambda_4) )
  \Phi=
  \tfrac{1}{4} (4 x^3+  \lambda_4 x^2 +  \lambda_3 x+  \lambda_2)   \Phi.
   \end{eqnarray*}
And the $  \Phi=  \Phi (u_0, u_1, u_2; (y, x) ) $ of the curve $y^2=4
x^5+  \lambda_4 x^4 +  \lambda_3 x^3+  \lambda_2x^2+  \lambda_1
x+  \lambda_0$ solves these equations for all $x$.

The most remarkable of the equations of Theorem   \ref{the-F} is the
balance of powers of the polynomials $  \gamma_{kl}$,  $  \beta_{kl}$
and of the ``spectral part'' --- the umbral derivative
$D_{k+l} (f (x) ) $:
  \begin{align*}
&  \mathrm{deg}_x  \gamma_{kl} (x,   \boldsymbol{ u})   \leqslant
g-1-  \mathrm{min} (k, l) ,    \\&  \mathrm{deg}_x  \beta_{kl} (x,   \boldsymbol{
u})   \leqslant 2g- (k+l) ,   \\
&  \mathrm{deg}_x D_{k+l} (f (x) ) =2g+1- (k+l) .
  \end{align*}

  \subsection{Solution of KdV equations
 by Kleinian functions}
The KdV system is the infinite hierarchy of differential equations
  \[
u_{t_{k}} =  \mathcal{ X}_{k}[u],
  \]
the first two are
  \[u_{t_{1}} =u_x ,   \quad  \text{and}  \quad u_{t_{2}} =
-  \tfrac12 (u_{xxx}-6u u_{x}) ,   \]
and the higher ones are defined by the relation
  \[
  \mathcal{ X}_{k+1}[u]=  \mathcal{ R}  \mathcal{ X}_{k}[u],
  \]
where $  \mathcal{ R}=-  \frac12  \partial_x^2+2u+u_x  \partial_x^{-1}$ is the Lenard's
recursion operator.

Identifying time variables $ (t_1=x,  t_2,   \ldots,  t_g)   \to (u_{g},
u_{g-1},   \ldots, u_1) $ we have

  \begin{prop}
The function $u=2  \wp_{gg} (  \boldsymbol{ u}) $  is a $g$--gap
solution of the KdV system.
  \end{prop}

  \begin{proof}
Really,  we have $u_x=  \partial_g 2  \wp_{gg}$  and
by   \eqref{wpgggi}
  \[u_{t_{2}}=  \partial_{g-1} 2  \wp_{gg}=
-  \wp_{ggggg}+12  \wp_{gg}  \wp_{ggg}.  \]
The action of $  \mathcal{ R}$
  \[   \partial_{g-i-1}2  \wp_{gg}=   \left[-
  \partial_{g}^2+8  \wp_{gg}   \right]   \wp_{gg, g-i}
+4  \wp_{g, g-i}  \wp_{ggg}  \]
is verified by   \eqref{wpgggi} and    \eqref{wp3}.

On the $g$--th step of recursion the ``times''  $u_i$ are
exhausted and the stationary equation \[ \mathcal{ X}_{g+1}[u]=0
\] appears.  A periodic solution of $g+1$ higher stationary
equation is a $g$--gap potential  (see   \cite{dmn76}) .
\end{proof}

\section*{Concluding remarks}

The Kleinian theory of hyperelliptic
Abelian functions as, the authors hope, this paper shows is an
important approach alternative to the generally adopted formalism
based directly on the multidimensional $\theta$--functions in
various branches of mathematical physics.  Still, a number of
remarkable properties of the Kleinian functions were left beyond
the scope of our paper. We give some instructive examples for the
case of genus two ${\boldsymbol u}=\{u_1, u_2\}$).

\begin{itemize} \item the
addition theorem \[ \frac{\sigma({\boldsymbol u}+{\boldsymbol
v})\sigma({\boldsymbol u}-{\boldsymbol v})}{\sigma^2({\boldsymbol
u})\sigma^2({\boldsymbol v})}= \wp_{22}({\boldsymbol
u})\wp_{12}({\boldsymbol v})-\wp_{12}({\boldsymbol u})\wp_{22}({\boldsymbol
v})+\wp_{11}({\boldsymbol v})-\wp_{11}({\boldsymbol u}) ,
\]

\item the equation, capable of being interpreted as the
Hirota bilinear relation:
\[
\left\{\frac13\Delta\Delta^T+\Delta^T\left(\begin{array}{ccc}
0&0&1\\0&-\frac12&0\\1&0&0\end{array}\right)\epsilon_{\eta,\eta}
\epsilon_{\eta,\eta}\cdot\epsilon_{\eta,eta}^T-(\xi-\eta)^4
\epsilon_{\eta,\xi}\epsilon_{\eta,\xi}^T\right\}\sigma({\boldsymbol u})
\sigma({\boldsymbol u}')\Big|_{{\boldsymbol u}'={\boldsymbol u}},
\]
is identically $0$, where
$\Delta^T=(\Delta_1^2,2\Delta_1\Delta_2,\Delta_2^2)$ with
$\Delta_i=\frac{\partial}{\partial u_i}-\frac{\partial}{\partial
u_i'}$ and also $\epsilon^T_{\xi,\eta}=(1,\eta+\xi,\eta\xi)$.
After evaluation the powers of parameters $\eta$ and $\xi$ are
replaced according to rules $\eta^k,\xi^k \to
\lambda_k\frac{k!(6-k)!}{6!}$ by the constants defining the curve.

\item  for the Kleinian $\sigma$--functions the operation is
defined
\[ \sigma(u_1,u_2)=\mathrm{exp}
\left\{\frac{u_2}{u_1} \sum_{k=1}^6  k
\lambda_k\frac{\partial}{\partial \lambda_{k-1}}\right\}
\sigma(u_1,0), \]
which resembles the function executed by vertex operators.
\end{itemize}
We give these formulas with reference to \cite{ba07}.

Another interesting problem is the reduction of
hyperelliptic $\wp$--functions to lower genera. In the case of
genus two it, happens according to the Weierstrass theorem when the
period matrix $\tau$ can be transformed to the form (see e.g.
 \cite{ba97,hu94})
\[
\tau=\left(\begin{array}{cc}\tau_{11}&\frac{1}{N}\\
\frac{1}{N}&\tau_{22}\end{array}\right),
\]
where so called {\it Picard number} $N>1$ is a positive integer.
The associated Kummer surface turns in this case to {\it
Pl\"ucker surface}. The reductions of the like were studied in
\cite{bbeim94} in order to single out elliptic potentials among
the finite gap ones. The problems of this kind were treated in
\cite{gw95, gw95a,gw95b}  by means of the spectral theory. We
remark that the formalism of Kleinian functions extremely
facilitates the related calculations and makes the solution more
descriptive.

These and other problems of hyperelliptic abelian functions will
be discussed in our forthcoming publications.

Concluding we emphasize, that the Kleinian construction of
the hyperelliptic Abelian functions  does not
exclude the theta functional realization but complements it, and
to the authors' experience the combination of the both approaches
makes the whole picture more complete and descriptive.

\section*{Acknowledgments}
The authors are grateful to S.P. Novikov for the attention and
stimulating discussions; we are also grateful to I.M. Krichever,
S.M. Natanson and A.P. Veselov for the valuable discussions.
Special thanks to G. Thieme for the help in the collecting
the classical German mathematical literature.

The research described in this publication was supported in
part by grants no. M3Z000  (VMB)   and no. U44000  (VZE)  from the
International Science Foundation and also the INTAS grant
no.  93-1324  (VZE and DVL),
and  grant no.  94-01-01444 from Russian Foundation of Fundamental
Researches.

\end{document}